\documentclass{article}  

\usepackage[left=1.25in,top=1.25in,right=1.25in,bottom=1.25in,head=1.25in]{geometry}

\usepackage{pdfpages}
\usepackage[section]{placeins}

\RequirePackage{amsthm,amsmath,amsfonts,amssymb}
\RequirePackage[authoryear]{natbib}
\RequirePackage[colorlinks,citecolor=blue,urlcolor=blue]{hyperref}
\RequirePackage{graphicx}

\renewcommand*{\thesection}{\arabic{section}.}
\renewcommand*{\thesubsection}{(\alph{subsection})}

\usepackage{verbatim}
\usepackage{dsfont}
\usepackage{float}
\usepackage{siunitx}
\usepackage{gensymb}

\usepackage[colorinlistoftodos]{todonotes}

\def\ym{\mathrm{ym}}
\def\m{\mathrm{m}}

\def\R{\mathbb{R}}
\newcommand{\argmin}{\mathop{\mathrm{argmin}}}

\usepackage{natbib}

\usepackage{authblk}
\title{Ocean Mover's Distance: Using Optimal Transport for
  Analyzing Oceanographic Data}

\author{
  \normalsize
  Sangwon Hyun$^{1}$\footnote{Author for correspondence; \texttt{sangwonh@ucsc.edu}},
  Aditya Mishra$^{2}$,
  Christopher L. Follett$^{3}$,
  Bror Jonsson$^{4}$,
  Gemma Kulk$^{4}$,
  Gael Forget$^{3}$,
  Marie-Fanny Racault$^{7}$,
  Thomas Jackson$^{4}$,
  Stephanie Dutkiewicz$^{3}$,
  Christian L. Müller$^{2,5,6}$,
  Jacob Bien$^{1}$
}

\date{
  \normalsize
  $^{1}$Data Sciences and Operations, University of Southern California, CA, USA\\
$^{2}$ Center for Computational Mathematics, Flatiron Institute, New York, NY, USA\\
$^{3}$ Department of Earth, Atmospheric and Planetary Sciences, Massachusetts Institute of Technology, Cambridge, MA, USA.\\
$^{4}$ Earth Observation Science and Applications, Plymouth Marine Laboratory, Plymouth, UK\\
$^{5}$ Department of Statistics, LMU München, Munich, Germany\\
$^{6}$ Institute of Computational Biology, Helmholtz Zentrum M\"unchen, Neuherberg, Germany\\
$^{7}$ School of Environmental Sciences, University of East Anglia, Norwich, UK\\
}

\begin{document}
\maketitle


\begin{abstract}

\small
Remote sensing observations from satellites and global biogeochemical models have combined to revolutionize the study of ocean biogeochemical cycling, but comparing the two data streams to each other and across time remains challenging due to the strong spatial-temporal structuring of the ocean. Here, we show that the Wasserstein distance provides a powerful metric for harnessing these
structured datasets for better marine ecosystem and climate
predictions. Wasserstein distance complements commonly used point-wise difference methods such as the root mean squared error, by quantifying differences in terms of spatial displacement in addition to magnitude. As a test case we consider Chlorophyll (a key indicator of phytoplankton biomass) in the North-East Pacific Ocean, obtained from model simulations, \textit{in situ} measurements, and satellite observations. We focus on two main applications: 1) Comparing model predictions with satellite observations, and 2) temporal evolution of Chlorophyll both seasonally and over longer time frames. Wasserstein distance successfully isolates temporal and depth variability and quantifies shifts in biogeochemical province boundaries. It also exposes relevant temporal trends in satellite Chlorophyll consistent with climate change predictions. Our study shows that optimal transport vectors underlying Wasserstein distance provide a novel visualization tool for testing models and better understanding temporal dynamics in the ocean.

Subjects: {\it Climatology, Oceanography}

Keywords: {\it Wasserstein distance, Earth mover's distance, Data-model
    comparison, Optimal Transport, Chlorophyll, Remote Sensing}

\end{abstract}

\maketitle

\section{Introduction} 

Understanding the differences between large spatiotemporal datasets is a common
task in oceanography.  Whether quantifying the agreement between the output of
an ocean simulation model
\cite{Dutkiewicz2015CapturingModel,Forget_2015} 
and \textit{in situ} measurement \cite{Moore2009,Jackson2017} or monitoring the changes
in the ocean across time \cite{Dutkiewicz2019OceanChange}, one needs a
meaningful notion of ``distance" between scalar fields defined across the ocean. 
We focus on the case in which the scalar field of interest represents the density or
concentration of a quantity over space.  It is most common to compare
images or data distributions using a ``pixel-by-pixel'' or pointwise difference
\cite{seegers2018performance, Forget_2015,Forget_2015b,Forget_2015c}; popular examples of such distances include
root-mean-squared error (RMSE) 
and mean absolute error. However, although easy to compute, pixel-wise comparisons 
may not fully account for the spatiotemporal nature of ocean 
data, which can exhibit complicated patterns composed of both global and local underlying
trends linked to shifting and evolving water mass bodies. 

These issues are well known and have led to the development of various normalized differences or ``cost functions'' which differentially weight
differences arising from deviations in quantity, location or from unresolved scales
(e.g. \cite{forget-wunsch-2007, Forget_2015, Forget_2015b}). Focusing on the probability
distribution over predefined regions (e.g., marine provinces, or
water masses) is one way to account for spatial errors. 
This method has been
used to examine, for example: the volumetric census of water masses
\cite{forget-2010,SPEER2013211}; relationships between primary production and
export \cite{cael2018can}; and the effects of mesoscale eddies
\cite{ashkezari-eddy}. Power-spectra further provide a useful basis for comparison as
a function of space and/or time scale (e.g. \cite{Forget_2015b}, \cite{McCaffrey-2015}). Despite these advances there remains a need for metrics which take into account pattern differences in a clear and interpretable way. This is especially true when evaluating the skill (or error) of ocean biogeochemical model simulations 
compared to other data sources such as satellite-derived measurements. Indeed, a recent summary paper \cite{ioccg-report} 
reports the need for a better measure of ocean Chlorophyll difference that goes beyond pixel-wise differences.
The reasons are many. Computer simulations may not be finely resolved  enough to capture meso-scale Chlorophyll patterns  (e.g. eddies) 
in time and space. However, such features will be captured \textit{in situ} and using satellites. Further, small spatial mismatches can result in large pixel-wise differences -- see Section 5.3.2 of \cite{ioccg-report}  -- which penalize models that are mechanistically correct for stochastic fluctuations. What we need is a metric which is easy to interpret, like RMSE, but for pattern differences.

In this paper, we explore the use of the Wasserstein distance \cite{villani2021topics}, which sometimes goes by the name {\em earth mover's distance} \cite{image-application-emd}. 
As that name suggests, Wasserstein distance measures the total amount of
''dirt''-moving that would be required to transform one mound of dirt
(representing a probability distribution) to make it equivalent to another mound
(a second probability distribution). The probability distributions in our
context are normalized versions of the scalar fields. Unlike pixel-by-pixel
distances, the Wasserstein distance incorporates the spatial structure of
discrepancies, making it particularly well-suited for the comparison of ocean datasets.
Wasserstein distance has been used in several other areas of geosciences. To list a few, 
it has been used to analyze particle distributions in the ocean \cite{Nooteboom2019}, for measuring error in temperature, precipitation, and sea ice projections \cite{Vissio2020}, for ocean data assimilation \cite{Tamang2020, Le2021}, for analyzing sea height images\cite{PapadakisThesis}, for ocean Synthetic Aperture Radar (SAR) segmentation \cite{Colin2021}, and for studying sea ice imagery \cite{Parno2019}. However, \cite{ioccg-report} makes clear that Wasserstein distance has not been thoroughly applied to the fundamental problem of model-to-data comparison and model-skill evaluation particularly in the context of ocean biogeochemical models and the representation of marine ecosystem structure and function. The goal of this paper is to carefully highlight the usefulness of Wasserstein distance in this context, as well as to show its usefulness in exploring time series of satellite maps. We focus on high-coverage Chlorophyll observations in the North Pacific
Subtropical Gyre \cite{Jackson2017}, and demonstrate how discrepancies between
model predictions and observed Chlorophyll can be interpreted in terms of a transport
field that when integrated over space yields a measure of distance in spatial
units.  We do this for the comparison of surface maps (see Section
3\ref{sec:lonlat}) and of depth profiles (see Section 3\ref{sec:depth}), which reveals long-term temporal trend and seasonality of satellite and model Chlorophyll maps in Section 3(a)\ref{sec:lonlat-nonclim}.

\begin{figure}
  \centering
  \makebox[1\linewidth]{
  \includegraphics[width = .33\linewidth]{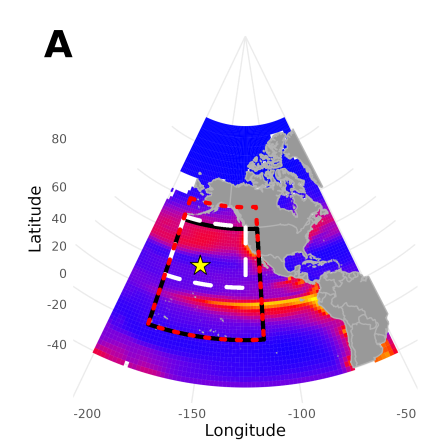}
  \includegraphics[width = .22\linewidth]{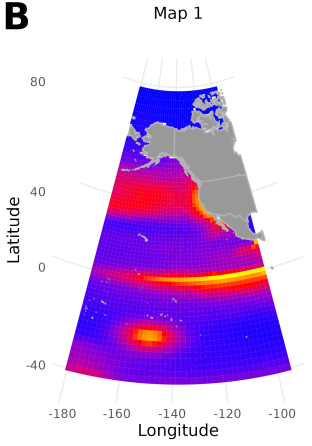}
  \includegraphics[width = .18\linewidth, trim={1.5cm 0 0 0},clip]{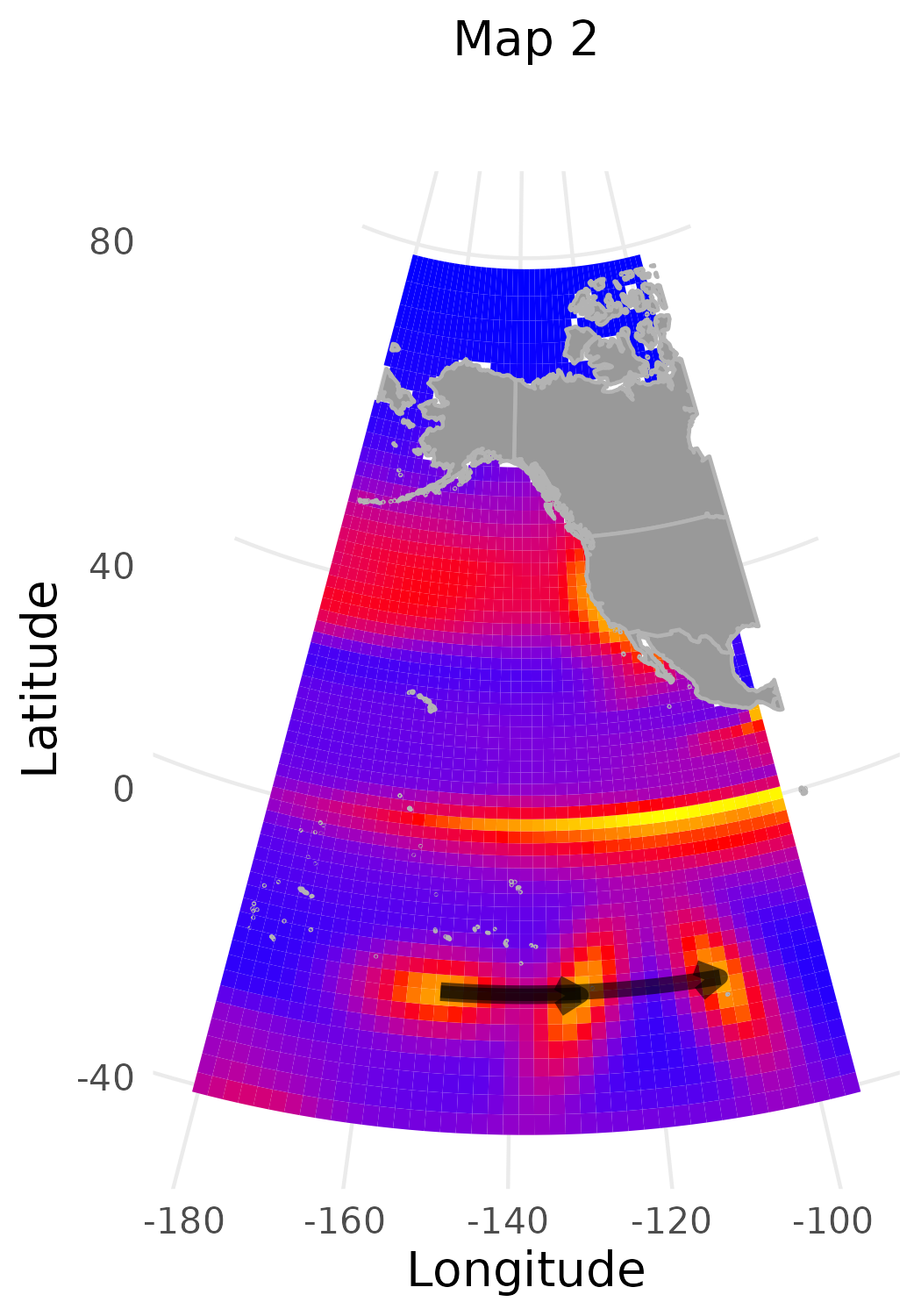}
  \includegraphics[width = .25\linewidth]{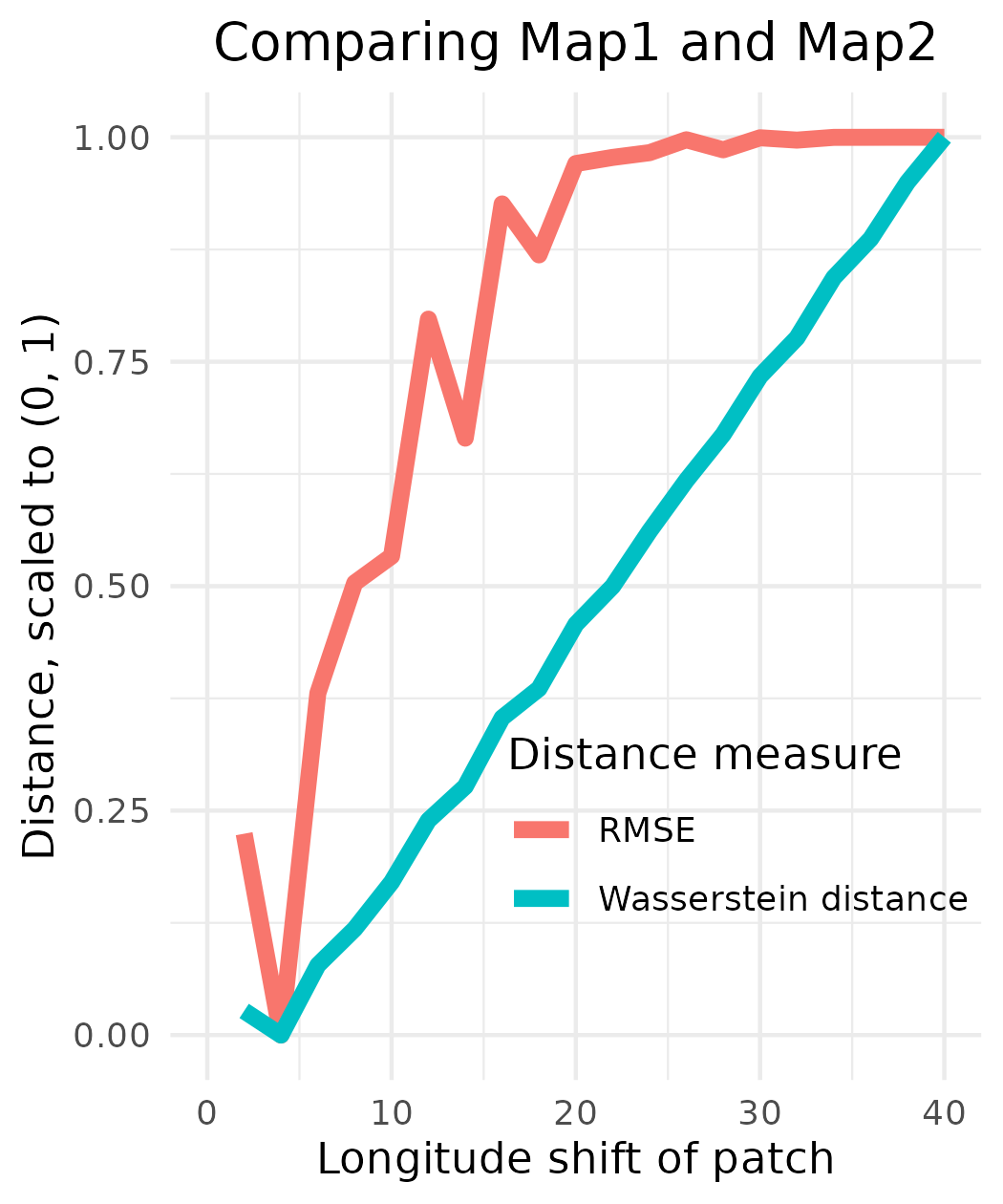} 
  }
  \caption{\it The left-most panel A shows a map of the study regions that are used
    for data analysis in this paper (solid lines for
    Section~3(a)\ref{sec:lonlat-clim}, dotted lines for
    3(a)\ref{sec:lonlat-nonclim}, and dashed lines for
    Section~3(a)\ref{sec:boundary-omd}); the star marker denotes the location
    of station ALOHA near Hawaii from which we obtain depth profiles of
    Chlorophyll to analyze in Section~3\ref{sec:depth}. In panel B, the first
    two figures show a toy example of two Chlorophyll maps both formed using
    simulated climatology data in January (from the ocean coupled
    physical-biogeochemical-optical model
    \cite{Dutkiewicz2015CapturingModel,Forget_2015}). One map was formed by
    adding an artificial patch of Chlorophyll to a longitude of --150. The other
    map was formed by shifting this patch to the east by up to 40 degrees
    longitude (while also rotating it). The right-most graph shows two different
    distance measures---root-mean-squared error (RMSE) and Wasserstein
    distance---between the two plots, while varying the amount of longitude
    shift of the patch. RSME plateaus after a shift of 20 degrees, while the
    Wasserstein distance is proportional to the amount of shift. }
  \label{fig:intro}
\end{figure}

To convey the intuitive appeal of the Wasserstein distance over pixel-wise distance measures, consider
the toy example in Figure \ref{fig:intro}, in which we imagine two surface maps
that are identical except for the location of an artificially inserted patch of
Chlorophyll south of the Equator. Physical processes like, for example, Rossby waves can generate such propagating patches.  The right panel shows how RMSE and Wasserstein distance quantify the difference between
the two surface maps as spatial shift of the patch increases.
RMSE quickly saturates: once the two patches have no spatial overlap, there is
no further change in the RMSE metric.  By contrast, the Wasserstein distance increases in an
approximately linear fashion. Indeed, the Wasserstein distance has units of distance and is directly related to the distance that the patch has moved.




In addition to its merit as a scalar distance, the Wasserstein distance also enables the visualization of the transport that would most efficiently (from the
perspective of a person moving the dirt) transform the first ocean map into the
second. For example, the rightmost panel of Figure~\ref{fig:climatology-mds}A
shows the optimal transport pattern between the two maps on the left (see
Section
3(a)\ref{sec:lonlat-clim}). 
These optimal transport patterns are \emph{not} to be interpreted as ``physical" transport
of the underlying quantity.  Still, these optimal transport patterns are useful
for understanding \textit{how} the data differ.  In this work, we consider two
primary types of comparison:
(1) 
comparing two different data sources measuring the same signal on a
spatiotemporal region or gridpoints; and (2) comparing the same data source at
different times. 
In both cases, visualizing the optimal transport can provide a scenario to 
elucidate the nature of the difference.  This can be particularly useful when
spatiotemporal differences are related to shifts in patterns that may not be well 
captured by pixel-wise comparisons.





With this paper, we aim to highlight the usefulness of studying ocean data using Wasserstein distance, 
which we show is particularly well-suited for evaluation of ocean biogeochemical models, 
among many other applications.
We compare satellite Chlorophyll
observations from the Eastern North Pacific Ocean and depth profiles from the
North Pacific Subtropical Gyre (NPSG) with their counterparts from a
biogeochemical model coupled to a state estimate of the ocean currents,
temperature, and salinity \cite{Forget_2015}. We show that the Wasserstein distance for
Chlorophyll between model and satellite data is large compared to the Wasserstein distance over the
seasonal cycle from satellite data or the model. We
further show how Wasserstein distance can be used to track changes in the transitional boundaries
between marine provinces over time \cite{Follett2021MovingPacific}. When reduced to 
this ``feature comparison'' we find that the model and satellite observations are in
relatively close agreement. 
Furthermore, applying a similar analysis to
the Chlorophyll depth profiles 
at Station ALOHA
\cite{karl1996hawaii,karl2014microbial}, 
discrepancy between model
outputs and {\em in situ} data is framed in terms of Chlorophyll shifts along the
depth dimension. Our numerical experiments allowed us to investigate whether the Wasserstein distance can effectively capture deviations in the ``Deep Chlorophyll Maximum'' between two Chlorophyll depth
profiles \cite{venrick1973deep,cullen1982deep,huisman2006reduced}. These results provide a path and
justification for using Wasserstein distance to analyze deviations in terms of pattern
displacements, and provide complementary information on magnitude differences.





\section{Material and Methods}

\subsection{Wasserstein Distance}
\label{sec:emd}

Consider two discrete probability distributions
$P = (P_i)_{i=1}^m, Q = (Q_j)_{j=1}^n,$ such that $P_i\ge0$ for all $i$,
$Q_j\ge 0$ for all $j$, and $ \sum_i P_i = \sum_j Q_j = 1$. In our context, $i=1,\ldots,m$ indexes a spatial partition of the region of ocean being studied into $m$ cells (and likewise for $j=1,\ldots,n$) and $P_i$ gives the proportion of the Chlorophyll (or any other positive quantity the scalar field is representing) in the region that is in cell $i$.  In the special case
that $i$ and $j$ index the same set of cells (such as $m=n$ pixels), one can define
pixel-wise distances such as the root-mean-squared error,
$\mathrm{RMSE}(P,Q) = \left( \frac{1}{n} \sum_{i} (P_i-Q_i)^2\right)^{1/2}$. If $P$ and $Q$
do not exist on the same coordinates, they need to be reconciled (processed) to
exist on the exact same cells in order to calculate RMSE. This
requirement is not shared by Wasserstein distance, which we describe next.

Wasserstein distance, which is also sometimes called earth mover's distance \cite{image-application-emd}, as discussed in
the introduction can be thought of as the total amount of ``dirt"-moving
required to transform a mound shaped like $P$ to a mound shaped like $Q$ when
one performs \textit{optimal transport} \cite{monge1781memoire,
  kantorovitch1958onthetranslocation, villani2021topics}, i.e. when one does
this earth moving in the most efficient fashion possible.
More precisely, the optimal transport between $P$ and $Q$ can be expressed
as solving the following linear program:
\begin{equation}
\hat f = \argmin_f {\sum _{i=1}^{m}\sum _{j=1}^{n}f_{i,j}d_{i,j}^2},
\text{ subject to } \begin{cases}
  \forall i,j:  f_{i,j} \ge 0  \\
  \forall i: \sum_{j}f_{i,j} = P_i \\ 
  \forall j: \sum_{i}f_{i,j} = Q_j, \\
\end{cases}
\label{eq:emd-definition}
\end{equation}
where $d_{i,j}$ is the \textit{base distance}  
between cell $i$ in $P$ and cell $j$ in $Q$. The optimization variable $f_{i,j}$ describes the amount of probability mass being transported from $i$ to $j$. The constraints encode that no mass is created or destroyed and that the net effect of the transport is to take $P$ to $Q$. The objective function is a weighted sum of squared distances (the square used in this paper makes this the "2-Wasserstein" distance), where the weights are given by the amount of probability mass being transported across all pairs of cells, $i$ and $j$. The optimum $\hat f$ is the
optimal transport between $P$ and $Q$, and the Wasserstein distance is defined to be the square root of the optimal value of this optimization problem:
$\mathrm{W_2}(P,Q)=\left(\sum_{i=1}^m \sum_{j=1}^n \hat f_{i,j} d_{i,j}^2\right)^{1/2}$.

Throughout, we use the \texttt{transport} R package \cite{transport-r-package},
which implements the algorithm in \cite{bonneel-omd-algorithm} in which each
discrete probability distribution first undergoes a multiscale transformation
and is decomposed into a weighted sum of Gaussian bases; then the optimal
transport problem is solved using a network simplex algorithm. This has $O(nm)$
computational complexity. Solving the optimal transport problem with a full
dense $d_{i,j}$ (base distance matrix as in equation \eqref{eq:emd-definition})
is prohibitively slow at moderate problem sizes like $n=m=10,000$. One
interesting and straightforward future improvement is to reduce the number of
transports needed by setting $d_{i,j}=\infty$ if $|i-j|>c$ for some threshold
$c$. Generally, there is a large literature on algorithms to calculate optimal
transport, of which we cite only a recent few. Among popular cutting-edge
algorithms are fast approximations in the Fourier space \cite{fourier-emd} and
in the wavelet space \cite{wavelet-emd}. Also popular is entropic regularization
\cite{sinkhorn-emd}, which is known as Sinkhorn distance. The most analogous
pre-existing application of Wasserstein distance is to digital image data, and has gained
popularity in recent years in the neural network literature
\cite{image-application-emd}. 

A distinctive feature of ocean applications (as
opposed to, for example, digital image applications), is that the base distance
$d_{i,j}$ cannot be taken to be Euclidean distance, especially
when the coordinates of the cells $i$ and $j$ are far apart. Instead, we take
the base distance to be the \textit{great circle} distance between the
(longitude, latitude) coordinates, which we compute using the \texttt{geodist}
package in R \cite{geodist-R}. Our work also offers fully reproducible code, 
via an \texttt{R} package named \texttt{omd} (\url{https://github.com/sangwon-hyun/omd}),
which could be used for other ocean studies.





\subsection{Multidimensional Scaling}

In our analysis, multidimensional scaling plots will be used to help us interpret distance matrices,
often highlighting seasonality and other relationships across time. Using Wasserstein distance as described in Section 2\ref{sec:emd}, we can take a collection of
maps and form a distance matrix $D \in \R^{N \times N}$, where $D_{ab}$ is the Wasserstein distance between normalized Chlorophyll maps $a$ and $b$. To help interpret the resulting
distance matrix, we visualize the maps' relationship to each other using
classical multidimensional scaling (classical MDS) \cite{mds-book, classical-mds}. This popular
data analysis technique seeks a configuration of points in the
two-dimensional plane whose Euclidean distances are close to those in an
inputted distance
matrix. 
That is, after computing the Wasserstein distance between all pairs of $N$
maps, the goal is to find a low-dimensional embedding, $z_1, \cdots, z_N \in \R^2$, for which
$\|z_a - z_b\|_2\approx D_{ab}$ for all maps $1\le a<b\le N$. An approximate closed-form
solution can be calculated using an eigen-decomposition of the doubly
centered matrix of squared distances. The details are provided in
Supplement Section~\ref{sec:mds-details}.

\subsection{Data}

The analysis is based on monthly Chlorophyll data from three different data
sources: derived from ocean-color remote sensing observations, the output from a
global biogeochemical circulation model, and integrated \textit{in situ}
observations. We use a subdomain of the model and remote sensing datasets
focused on a latitude-longitude rectangle in the Pacific Ocean directly above---and including---Hawaii. The region is centered around about 20 degrees latitude
and $-155$ degrees longitude and captures interesting geographic variability in
the ocean. To the south of this region is the North Pacific Subtropical Gyre
(low latitude, dominated by warm, more saline water) and to the north is the
Subpolar Gyre (high latitude, low-temperature, low-salinity, nutrient-rich
water). The region between these two gyres is the North-Pacific Transition Zone
(NPTZ) with a strong gradient in Chlorophyll, as can be seen in the remote
sensing observations and in the model output (Figure~\ref{fig:climatology-mds}A,
left panels). We also focus on data directly from a fixed location near the
south of this region, Station ALOHA ($22.75$ degrees latitude and $-158$ degrees
longitude) \cite{aloha-data}. Throughout, we exclude Chlorophyll data near the
coastline where both satellite measurements and numerical models have known
irregularities. Each dataset is described in some detail next.




 \subsubsection{CBIOMES-global Model Output}

Model data is based on output from a coupled physical-biogeochemical-optical model, modified for the Simons Collaboration on Computational Biogeochemical Modeling of Marine Ecosystems (CBIOMES) project. The CBIOMES-global model simulates the period from 1992-2011 \cite{gael_forget_2018_1343303}.

The model's physical component is derived from the Estimating the Circulation and Climate of the Ocean project (ECCO), version 4 (ECCOv4) \cite{Forget_2015,  Forget_2015c, Forget_2015b}.  ECCOv4 uses a ``least-squares with Lagrangian multipliers" method to get internal model parameters, initial, and boundary conditions that minimize the discrepancy between global observational data streams of satellite and {\em in situ} data. The end product is a global three-dimensional configuration state estimate, at a horizontal resolution of 1 degree and with depth ranging from 10 m at the surface to 500 m at depth (see \cite{Forget_2015} for details). 

The biogeochemical/ecosystem component is from the MIT Darwin Project and follows that of \cite{Dutkiewicz2021a}. The model data we use in this paper is the aggregated Chlorophyll-a across all phytoplankton groups simulated from this ecosystem model, made into monthly averages. The amount of Chlorophyll in each of the $35$ phytoplankton types varies based on light, nutrients and temperature \cite{Geider1998ATemperature}. The $35$ phytoplankton types are from from several biogeochemical functional groups such as pico-phytoplankton, silicifying Diatoms, calcifying coccolithophores, mixotrophs that photosynthesize and graze, and nitrogen fixing diazotrophs, with sizes that span from 0.6 to 228 $\mu$m equivalent spherical diameter (ESD). The model incorporates various interactions with chemical factors (e.g. carbon, phosphorus, nitrogen, silica, iron, oxygen) and with other species (e.g. grazing by zooplankton). See \cite{Dutkiewicz2021a} for full details. Hereon, we will simply refer to this data as model data.

\subsubsection{Remote Sensing Data}

  Remote sensing (or satellite-derived) data is based on version 3.0 of the European Space Agency Ocean Colour Climate Change Initiative (OC-CCI) \cite{Melin2017, Sathyendranath2019,  shubha:2020},  a blended Level 4 Chlorophyll product with a spatial resolution of 4\,km. The OC--CCI V5.0 combines data from five independent ocean-colour sensors to produce merged, climate-quality observations of Chlorophyll concentration.  The sensors include the Sea-viewing Wide-Field-of-view Sensor (SeaWiFS), the Aqua MOderate-resolution Imaging Spectroradiometer (MODIS-Aqua), the MEdium spectralResolution Imaging Spectrometer (MERIS), the Suomo-NPP Visible InfraredImaging Radiometer Suite (NPP-VIIRS), and the Sentinel 3A Ocean and Land Colour Instrument (OLCI). 
  These data sources are algorithmically merged and processed (see more details of this processing in \cite{Jackson2017, Sathyendranath2019}), then downscaled to the same spatial grid as model data at the monthly time resolution.

\subsubsection{{\em In-Situ} Data from Station ALOHA} We additionally consider shipboard measured Chlorophyll-a from  Station ALOHA (22\degree{}45'N, 158\degree{}00'W). The dataset (obtained from the Simons Collaborative Marine Atlas Project (CMAP), originally sourced from \url{https://hahana.soest.hawaii.edu/hot/dataaccess.html}) contains concentrations of Chlorophyll collected using a CTD fluorescence sensor. There are $28,583$ observations measured between 1988-10-3 to 2016-11-27, in the depth range between 0 and 200 meters. This data was downloaded directly from \cite{cmap4r}, an \texttt{R} package for accessing the CMAP database. 


In Section~3\ref{sec:depth}, we compare \textit{depth profiles} (measurements
over depth) of {\em in situ} data and model data using Wasserstein distance. {\em In situ} data is sampled irregularly in time, while Darwin data is complete in space and time. In order to compile the two datasets at matching locations in space and time, we \textit{colocalize} the model data, by taking averages of the Chlorophyll measurements in a certain space-time vicinity
($\pm 2$ days and $\pm 5$ meters) of each time point of the {\em in situ} data. Panel B of Figure~\ref{fig:depth} shows
the Chlorophyll data from the two sources. Each depth profile is normalized by
dividing by the total so that the sum is 1 prior to calculating Wasserstein distance, as done for the maps.

\section{Results}
\label{sec:results}

\subsection{Geographical and Temporal Analysis of Chlorophyll Data}
\label{sec:lonlat}

In this section, we show several different data applications of Wasserstein
distance to the ocean setting, each highlighting a different aspect of ocean
data comparisons. First, in Section~3(a)\ref{sec:lonlat-clim} we consider the
climatological seasonal changes in Chlorophyll patterns in both satellite and
model, and we also perform direct model-satellite comparisons. Here, "climatological" refers to being based on the twelve average monthly Chlorophyll levels (averaging from 1998 to 2006). Next, in Section~\ref{sec:lonlat-nonclim} we
consider the full time series of monthly averages from 1998 to 2006 and focus on
using Wasserstein distance to explore change in Chlorophyll patterns over that time
period. Finally, in Section~\ref{sec:boundary-omd} we use a smaller longitude-latitude
rectangle in the North Pacific Transition Zone, and base comparisons on estimated boundaries between regions instead of on the original Chlorophyll concentrations.


\subsubsection{Climatology Chlorophyll Data}
\label{sec:lonlat-clim}

\begin{figure}[htb!]
\centering
  \includegraphics[width = \linewidth]{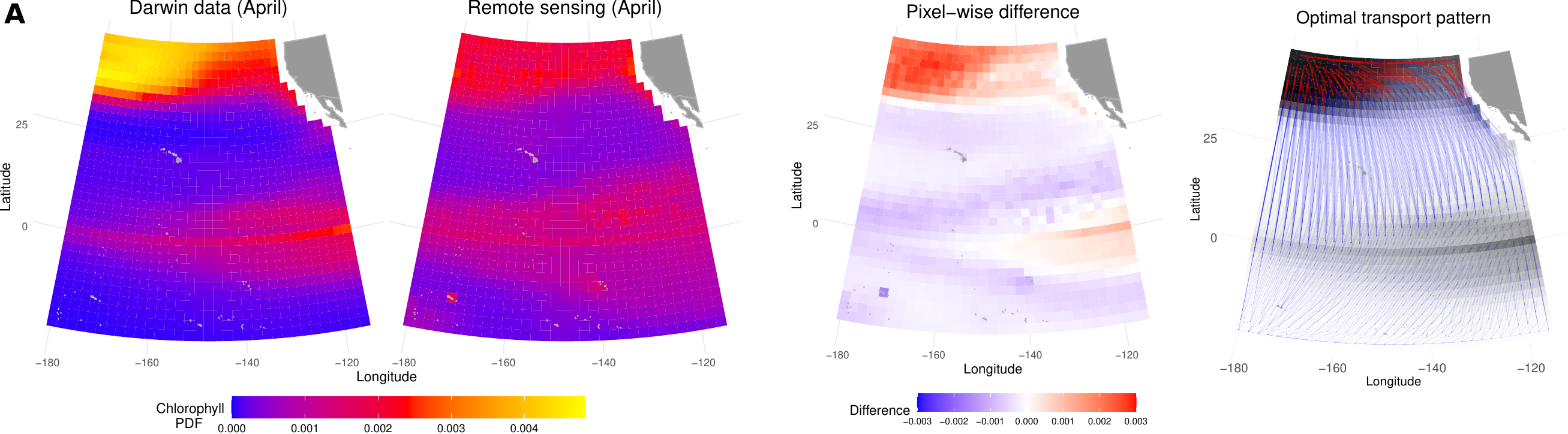}
  \hspace{-5mm}
  \includegraphics[width = .45\linewidth]{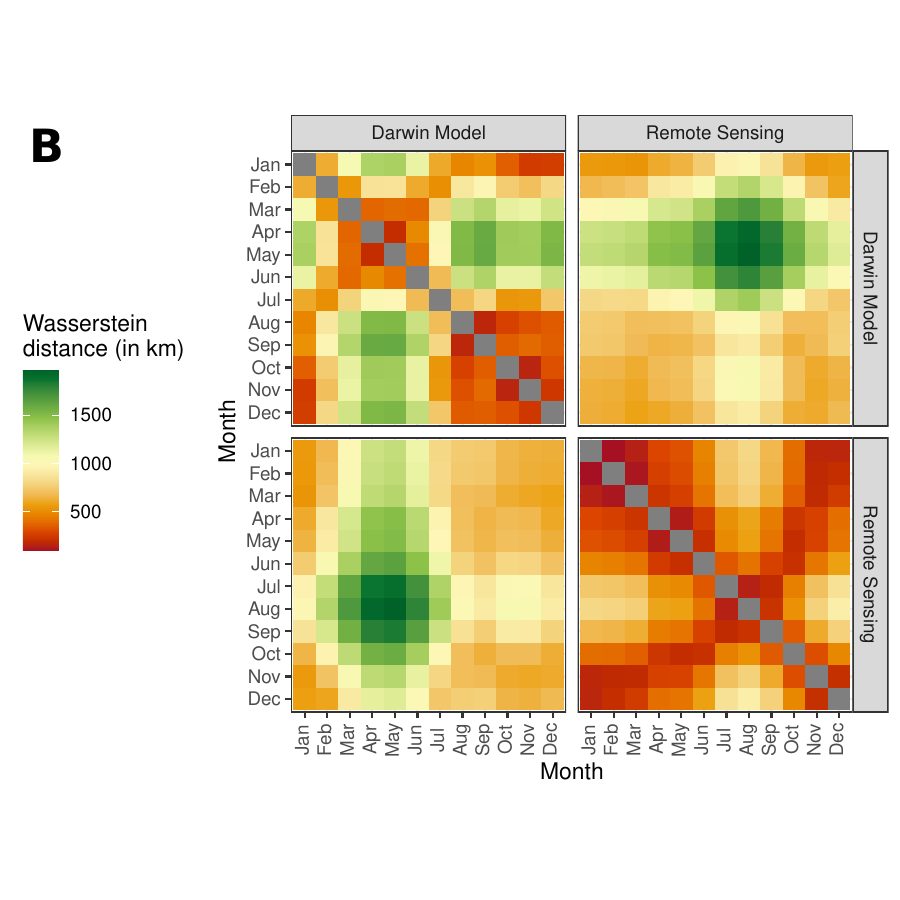} 
  \includegraphics[width = .47\linewidth]{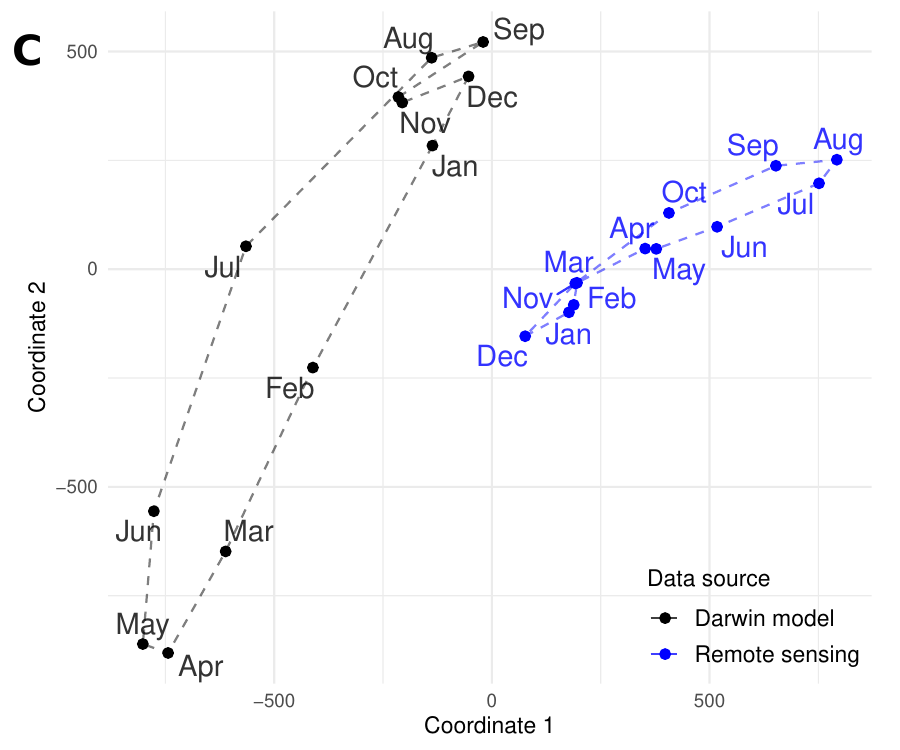}
  \caption{ \it Geographical analysis of Chlorophyll data. Panel A shows a comparison
    of April's climatology Chlorophyll maps from two data sources (two left
    maps) using optimal transport. The first two maps are measurements on a two-dimensional
    grid in which each grid cell measurement can be thought of as a pixel
    intensity in a digital image. The values have been normalized to sum to 1 in
    each map.  The third map in Panel A shows the pixel-wise difference (which is the basis for root-mean-squared error---RMSE) of the two left maps.  The rightmost
    map shows the estimated optimal transports (which is the basis for Wasserstein distance),
    with transparent blue arrows and opaque red lines showing the bottom 90\%
    and top 10\%  of all the masses, respectively. This mass transfer
    plot shows that the major shift of Chlorophyll probability mass from the
    concentrated upper left corner is dispersed in a south- and east-bound
    direction with a particular trend. Panel B shows a summary of all pairwise
    Wasserstein distances from the $24$ maps---twelve months of climatology Chlorophyll maps,
    from the two data sources (model and remote sensing), as a $24 \times 24$ distance
    matrix. Panel C shows a classical multidimensional scaling (MDS) performed
    on this data. Three notable observations can be made: (1) model data is more
    variable than remote sensing data, (2) there is a clear separation between
    model and remote sensing data, and (3) the connecting dashed line between
    adjacent months in each data source shows an annual seasonality. This is
    further explored in \ref{fig:nonclim-trend} and
    Figures~S\ref{fig:nonclim-mds} and by analyzing data from each year.}
  \label{fig:climatology-mds}
\end{figure}

Our first comparison is between the two climatology data sources---remote
sensing and model data. The third panel in Figure~\ref{fig:climatology-mds}A shows the pixel-wise difference, and portrays both large positive deviations in the northern region and smaller ones in a wider region near the equator. The rightmost panel shows an example of the optimal transport
pattern from comparing climatology remote sensing data and model data in April. Optimal transport is visualized as blue transparent arrows, and those corresponding to the top $10\%$ are highlighted in bold red. Both plots indicate that the model and remote sensing data differ the most in the northern region, while optimal transport additionally shows a southbound shift in patterns across the whole domain.


Next, we form a $24$-by-$24$ distance matrix $D = (D_{a,b})_{a,b}$, shown in
Panel B of Figure~\ref{fig:climatology-mds}, from the ${24 \choose 2}$ unique
pairwise Wasserstein distances between Chlorophyll maps $a$ and $b$ (ranging over all 12 months
and both data sources). This shows interesting seasonal changes in Chlorophyll
patterns within each of the data sources. For instance, the Wasserstein distances in a given row
(or column) in the top left panel (model) or bottom right panel (satellite) form
a unimodal curve when plotted as a 1-dimensional time series.  Also, the Wasserstein distances
between monthly remote sensing data in the top-left quadrant have much larger values
than the Wasserstein distances between monthly model data in the bottom-right quadrant,
meaning that patterns of Chlorophyll shift geographically more in the Darwin
model compared to the remote sensing data. The twelve Wasserstein distances between the 
two sources in each calendar month are shown in the diagonal values of the
upper-right and lower-left quadrants and have large values compared to (i)
the distances between any two months and (ii) the distances between adjacent months
in either data source.



We further summarize the distance matrix $D$ with a classical MDS plot (Panel C of Figure~\ref{fig:climatology-mds}), projecting the $24$ Chlorophyll maps onto
a 2-dimensional plot. This MDS plot again shows that model data has higher
variability than the remote sensing data. It also shows a clear separation between
the two data sources. The line connecting the data sources
shows a closed loop within each source, which shows seasonality according to
time of year. A careful look reveals that the seasonality pattern is different
for the two data sources---the distance between the three months (August
through October) and (December through January) is smaller in model data than in the
remote sensing data.

\subsubsection{Interannual Variability and Long-term Trends}
\label{sec:lonlat-nonclim}





\begin{figure}[htb!]
  \includegraphics[width = \linewidth]{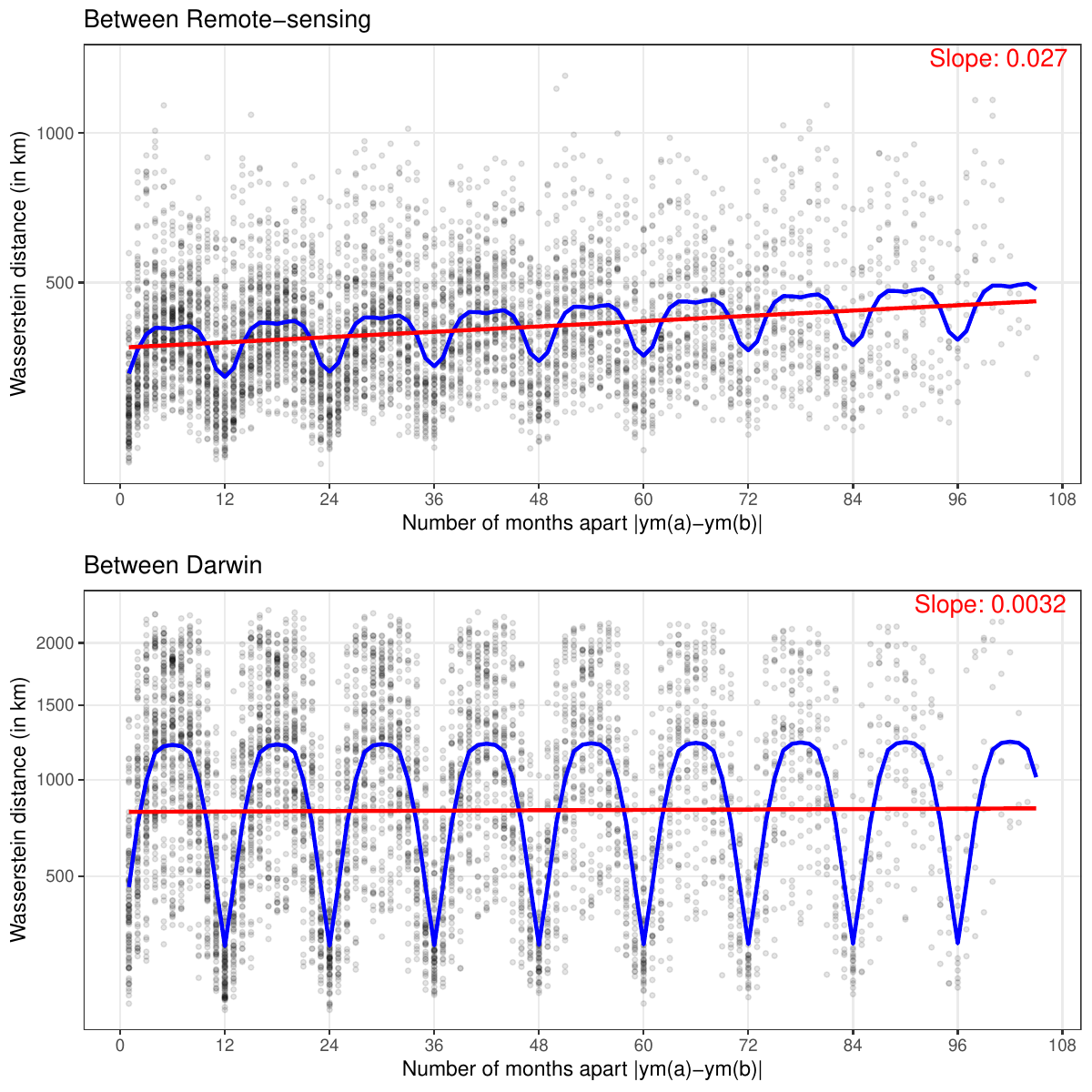} 
  \caption{\it Wasserstein distance between time-resolved Chlorophyll data from different months 
    (between March 1998 to December 2006) for the remote sensing (Panel A) and model (Panel B) data, arranged so that the $x$-axis shows how
    many months apart the two Chlorophyll maps are, and the $y$-axis is the Wasserstein distance
    (which uses square-root scaling). The blue line is fitted using a regression model that assumes a linear trend together with a regular seasonal pattern, and the red line shows the linear trend excluding the
    seasonal component. The slope of the red line for the remote sensing data is 
    roughly $8.4$ times
    larger than for the model data -- both slope values are shown in the top-right of each panel. Note, the red line is linear in $\sqrt{D_{ab}}$, and only appears linear here because the slope coefficient is very small in size. 
    }
  \label{fig:nonclim-trend}
\end{figure}

We expand the analysis by using time-resolved data based on monthly
averages of model and remote sensing data in all months available from 1998 to
2006. An MDS analysis leads to similar conclusions as those from the climatology
data (see Supplemental Section~\ref{sec:nonclim-mds} for a detailed analysis).
Next, Figure~\ref{fig:nonclim-trend} plots the Wasserstein distance between pairs of maps from
within a single source (model or remote sensing) as a function of the number
of months they are apart. The blue line shows a regression mean that explicitly
models annual seasonality, and the red line is the linear trend without the
seasonality. The regression model predicts $\sqrt{D_{ab}}$ between year-month $a$ and $b$, using two types of predictors: (i) the number of months apart $|\ym(a)-\ym(b)|$ that
includes year information and the (ii) number of \textit{calendar months} apart
if one ignores the years, i.e. $|\m(a)-\m(b)|\in\{0,\ldots,6\}$. The predictor in (ii) is an explicit accounting for differences in the time of year. In particular,
the fitted model for $\sqrt{D_{ab}}$ shown by the blue line is given by
\begin{equation}
  \hat\beta_0 + \hat\beta_1\cdot|\ym(a) - \ym(b)| + \sum_{k=0}^6 \hat\beta_{2,k}
  \mathds{1}(|\m(a)-\m(b)| = k),
  \label{eq:regression}
\end{equation}
where $\sum_{k=0}^6\hat\beta_{2,k}=0$. The red line is simply the first two terms of the above expression.
The undulating blue line indicates the larger seasonal variability in Chlorophyll patterns in the model relative to remote sensing data noted in Section~3(a)\ref{sec:lonlat-clim}.
The slope of the red line, $\hat\beta_1$, is positive for
remote sensing and 8.5 times that of the model data. 
Indeed, the upward trend of the red line for the remote sensing data is visibly
much more apparent than that for the model data.  This suggests that the
Chlorophyll maps in the remote sensing data are getting increasingly more
different from each other (i.e. there is a trend in the Chlorophyll patterns) in
a way that is not reflected in the model. This is further supported by
Figure~S\ref{fig:nonclim-trend-long} that shows a sustained trend in the remote
sensing data over a longer time period (1996-2020), as well as by the MDS plots
in Figure~S\ref{fig:nonclim-trend-long-from-mds}. Using RMSE instead of Wasserstein distance in
Figure~S\ref{fig:nonclim-trend-rmse}, the increasing trend is weaker but still
present, and about 2 times larger in remote sensing data than in model data.


\begin{figure}[htb!]
  \centering
  \includegraphics[width = .9\linewidth]{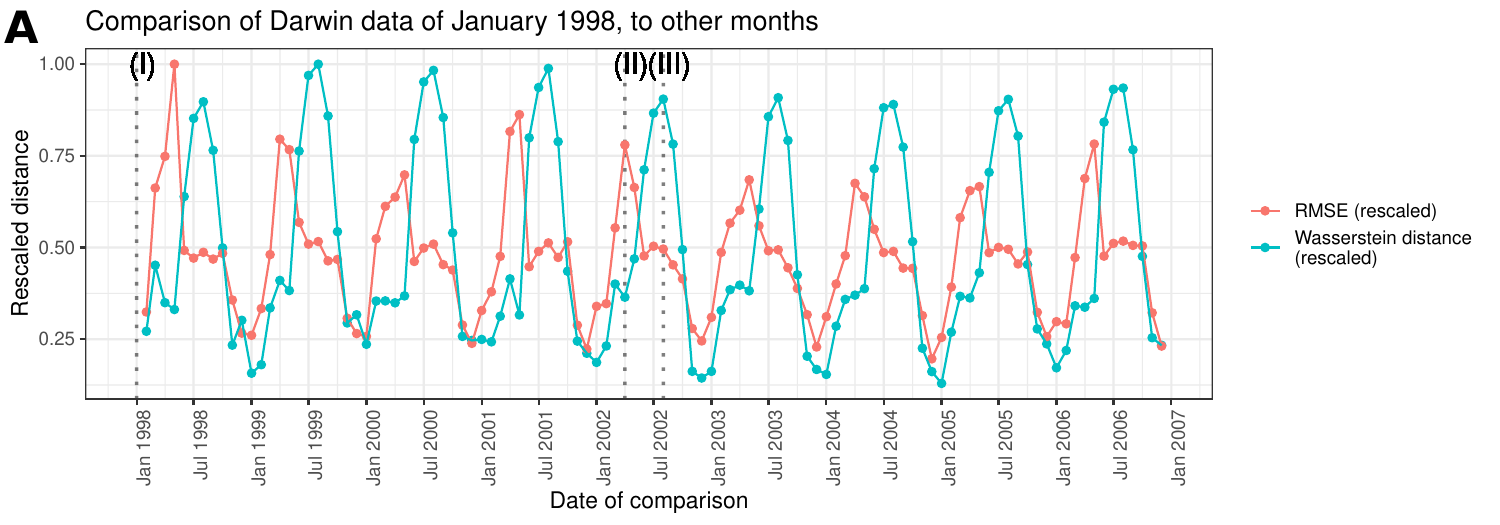}  
  \includegraphics[width = .9\linewidth]{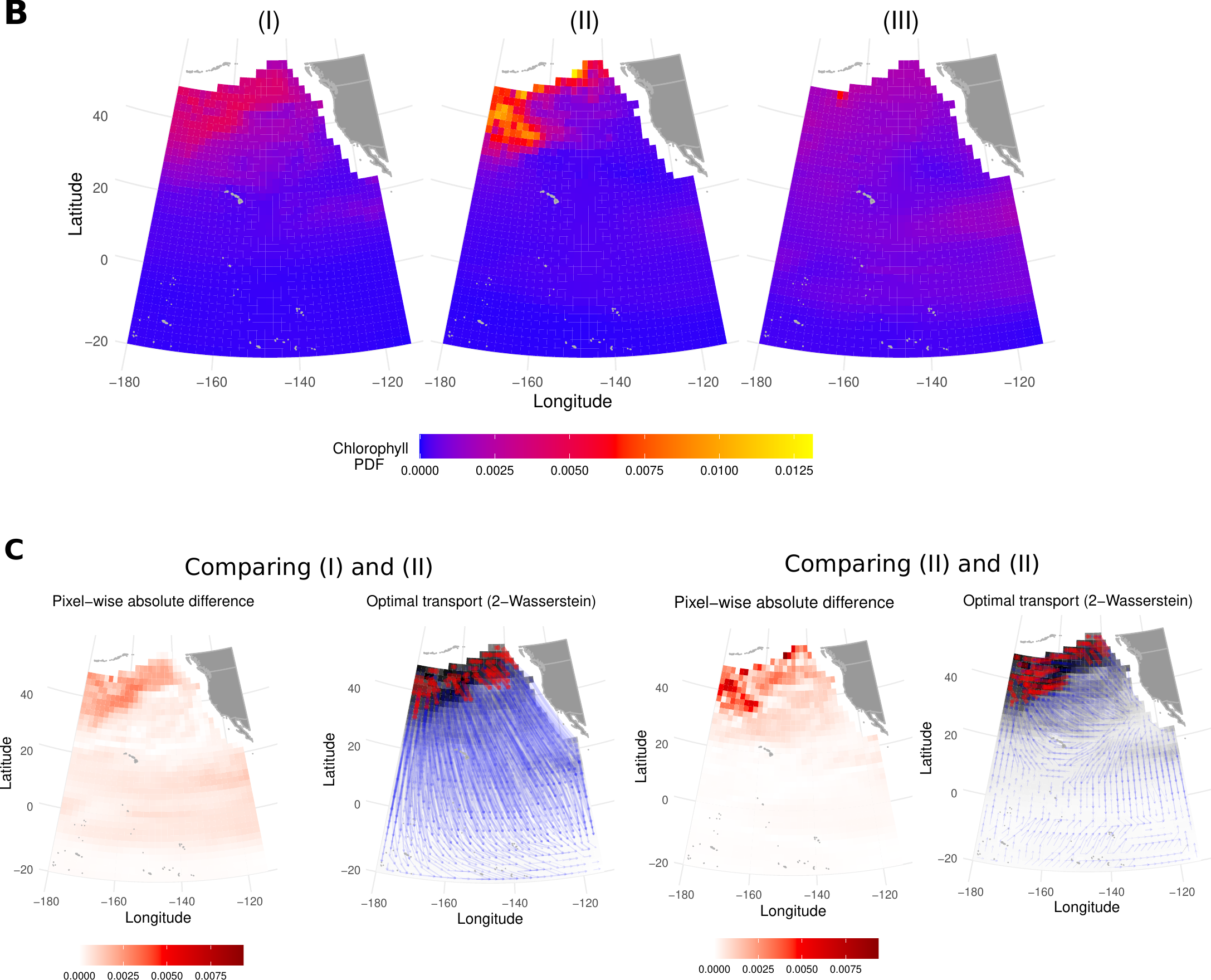}  
  \caption{ \small \it Comparison between the interpretation of time series data using Wasserstein distance and root-mean-squared error (RMSE). Panel A shows the distance between January 1998 model data (I) to
    other months' model data, measured by Wasserstein distance and RMSE, with distances normalized to range from 0 to 1. Panel B shows the three maps. Noticing that the seasonal cycle and annual
    peak of Wasserstein distance is different in the two sources, we focused on two months---April
    2002 (II) when RMSE peaks and Aug 2002 (III) when Wasserstein distance peaks. (Grey vertical dotted lines are drawn at dates (I), (II), and (III) for emphasis.) RMSE
    measures January to be more different from April than it is from August, while Wasserstein distance
   measured the opposite. In panel C, the optimal transport between (I)
    and (II) is mostly short shifts locally in the north, while the pixel-wise
    difference is overly pronounced due to a few large differences in the
    northern coastal region. On the other hand, the optimal transport between (I) and
    (III) includes two types of shifts---those that are local to the northern
    region, and sizeable equator-bound shifts. The pixel-wise difference does
    not capture the latter. Note, only half of the arrows are shown in the
    optimal transport plots for visual clarity. The year of $2002$ was chosen randomly, and the same analysis using another comparison year shows similar conclusions.
    }
  \label{fig:nonclim-zoomin}
\end{figure}

Lastly, Figure~\ref{fig:nonclim-zoomin} highlights a stark contrast between Wasserstein distance and RMSE. The lines plotted
in Panel A show the distance from model data in January 1998 to all other
months of model data in our date range, measured in two ways (Wasserstein distance and
RMSE). Both have regular seasonality, but the Wasserstein distance curve peaks in the summer (around
August) of each year, while the RMSE curve peaks in the early Spring (around April). We
focus on three months---shown as January 1998 (I), April 2002 (II), and August
2002 (III) in Panel A---and note that the domain of calculations have been extended 
further northward as compared with Figure~\ref{fig:climatology-mds}.

In Panel B comparing (I) and (II), we see that the RMSE is relatively high due
to a few large mismatches in the coastal region, while the Wasserstein distance in this
comparison is relatively small because only local shifts exist in the
North. On the other hand, Panel C comparing (I) and (II) shows that Wasserstein distance is
appropriately large; the rightmost figure shows how optimal transport captures
many global south-bound shifts in probability mass to the equatorial
region. Pixel-wise difference (third figure from the left) fails to capture this
visibly large pattern difference, and RMSE is measured to be smaller than from
the comparison in Panel B. This demonstrates how Wasserstein distance can be an improvement over RMSE in
quantifying such differences between maps.

\subsubsection{Comparing Ocean Provinces}
\label{sec:boundary-omd}


Sometimes, rather than comparing the scalar fields directly, we may be more
interested in comparing a scientifically relevant derived feature of the fields.
For example, 
one may algorithmically segment the ocean into 
cohesive regions---"provinces"---based on underlying differences in one or more fields (e.g. \cite{kavanaugh-hierarchical,irwin2008,Sonnewald2020b,Wust2020b}. 

\begin{figure}[htb!]
\centering
  \includegraphics[width = .25\linewidth]{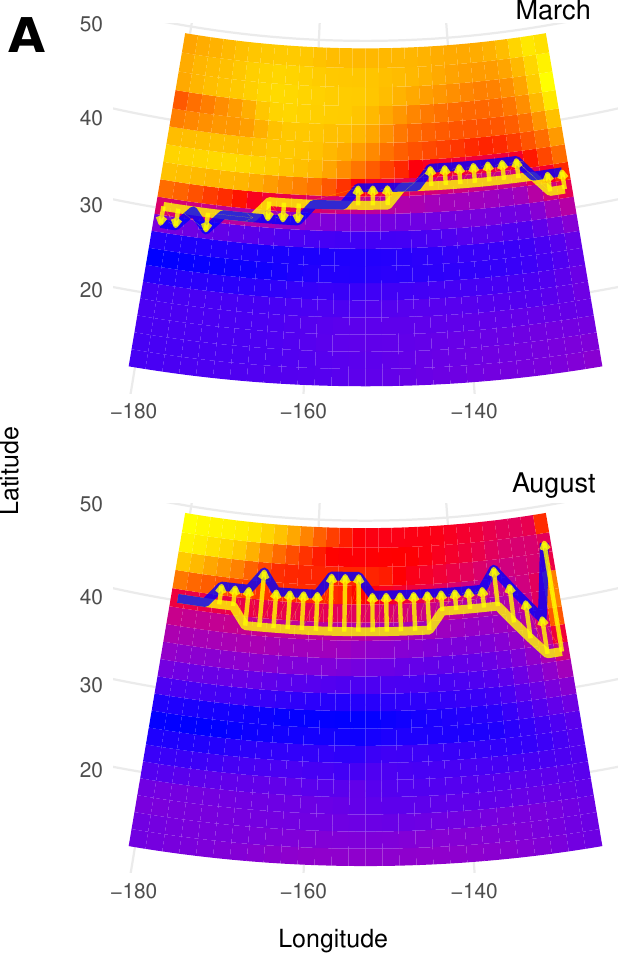} 
  \includegraphics[width = .38\linewidth]{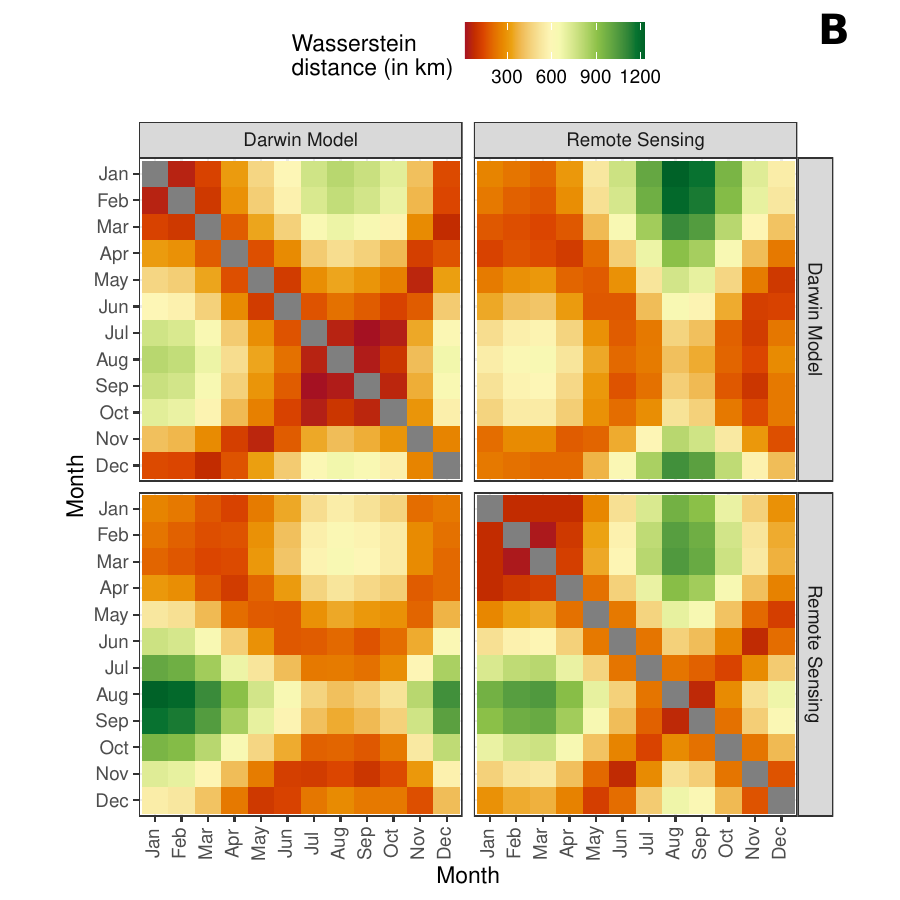} 
  \includegraphics[width = .3\linewidth]{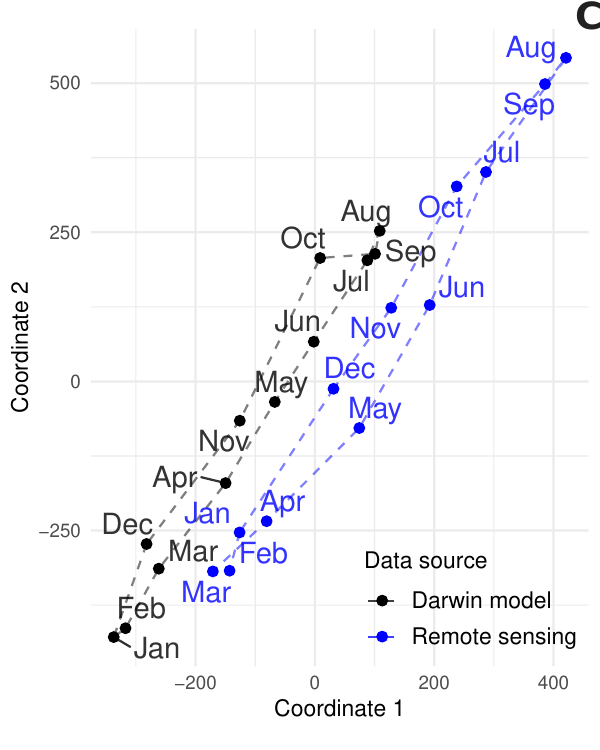} 
  \caption{\it Comparison of ocean provinces using Wasserstein distance ($W_2$). Panel A shows an example of
    the application of Wasserstein distance on cluster boundaries for March and August based on
    Chlorophyll climatology data (the full set of plots from all months are
    provided in Figure~S\ref{fig:boundary-all}). The plots show province
    boundaries estimated from remote sensing (blue line) and model (yellow
    line) data, overlaid on model Chlorophyll data shown as heatmaps. The next
    two panels show summaries of all pairwise Wasserstein distances from the $24$ maps of
    estimated cluster boundaries (for the twelve months of climatology
    Chlorophyll maps from the two sources) in the
    same style as Figure~\ref{fig:climatology-mds}. Panel B shows a $24 \times 24$
    distance matrix, and Panel C shows a classical multidimensional scaling (MDS) performed on this data. The
    distance between the two data sources in the same month is small and the
    seasonal dynamic shown by the lines is similar in the two data sources. This
    shows that, despite the large between-source distance between Chlorophyll
    maps in \ref{fig:climatology-mds}, one important aspect---the estimated
    boundary between the two bodies of water (the North Pacific Transition Zone and the Subtropical Gyre)---is similar between the two data sources.
      }
  \label{fig:boundary}
\end{figure}

We show here how Wasserstein distance can be used to evaluate how different the boundaries are of such provinces when determined from different datasets or algorithms. Here, we apply a clustering
algorithm (K-means clustering) to two Chlorophyll maps---one from remote sensing and the other from
the model---to estimate two different spatial provinces of
Chlorophyll. In our study region, this province boundary occurs in the North Pacific Transition Zone and is often referred to as the Transition Zone Chlorophyll Front (TZCF) \cite{Polovina2000SensorsPacific,Follett2021MovingPacific}. We demonstrate in this section how to use Wasserstein distance to flexibly
measure the difference between ocean provinces, by measuring how much transport
is needed to move the boundaries of one set of provinces (based on model data)
to make them equivalent to that of an alternative definition of provinces
(based on remote sensing data). 
Given a partition of the ocean, we can extract a binary scalar field that is 0 inside the provinces and equal to a nonzero constant along the discretized boundaries between regions. Given two such binary scalar fields, we can then apply Wasserstein distance.
An example is shown in Panel A of Figure~\ref{fig:boundary} for the March and August Chlorophyll climatologies, where the estimated boundary is shown as yellow (model) and blue (remote sensing) lines.


It is interesting to compare the distance matrices (Panels B) and the MDS plots (Panels C) in Figure~\ref{fig:boundary} and Figure~\ref{fig:climatology-mds}, which was formed by applying Wasserstein distance to the Chlorophyll field itself. When performing Wasserstein distance on the boundaries, the MDS plot in Figure~\ref{fig:boundary} shows little between-source difference (compared to within-source seasonal variability), with the months from the two data sources lining up with each other. By contrast, the MDS plot of Figure~\ref{fig:climatology-mds} showed a larger degree of between-source variability. 
In other words, despite the relatively large between-source distance between Chlorophyll maps, we see that in terms of one important aspect---the estimated boundary between the regions---the two data sources agree rather well. Putting this in the context of data source comparison, boundary comparison show a much better connection between the model and remote sensing data than the Chlorophyll fields themselves, suggesting the model captures the overarching patterns and controls although not the exact locations and more detailed patterns.

\begin{figure}[htb!]
  \includegraphics[width = \linewidth,page = 5]{figures/OMD_depthAnalysis.pdf}
    \includegraphics[width = \linewidth]{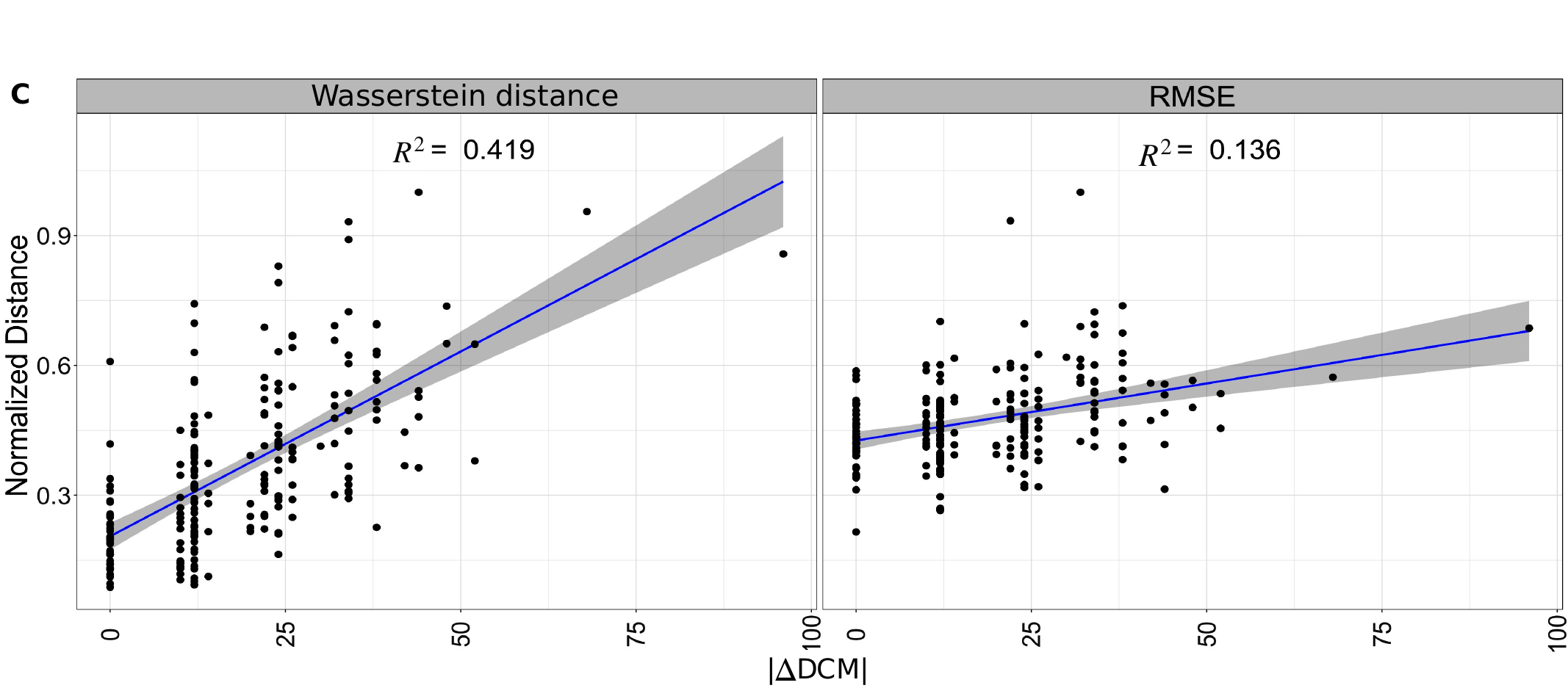}
  \caption{\it Comparing depth profiles of Chlorophyll from to data sources. Panel A and B show depth profiles of Chlorophyll from two data sources---model and {\em in situ}---with an example of a single depth profile for 2013-09-15 given in Panel A and all depth profiles between October 1988 and November 2016 (n = 226) given in Panel B. Each
    vertical slice (a single 1-dimensional histogram of Chlorophyll distribution
    from each data source) at overlapping time points can be compared using
    Wasserstein distance ($W_2$). Panel C shows the effectiveness of the two distance measures,
    root-mean-square error (RMSE) and Wasserstein distance ($y$-axis), in capturing the difference in the deep Chlorophyll
    maximum (DCM: $x$-axis) recorded from the model and at Station ALOHA ({\em in situ}) at shared
    dates. DCM refers to the region below the ocean surface where the maximum
    concentration of Chlorophyll is observed. The higher $R^2$ demonstrates that Wasserstein distance is better able to capture the variability in the difference in the DCM than RMSE.
    }
  \label{fig:depth}
\end{figure}

\subsection{Comparing Depth Profiles of Chlorophyll}
\label{sec:depth}

In this section, we use Wasserstein distance to compare Chlorophyll depth profiles at Station
ALOHA using two different data sources ({\em in situ} and model).  In the vertical
profile of Chlorophyll, a \textit{Deep Chlorophyll Maximum} (DCM) (sometimes
also referred to as a Subsurface Chlorophyll Maximum, SCM \cite{Anderson1969})
is observed as a pronounced peak at depth (generally below the first optical
depth) (Figure~\ref{fig:depth}). A DCM develops under stratified conditions
\cite{Estrada1993} at the point of cross-over between two conditions that limit
phytoplankton growth. Surface waters are light-rich and nutrient-limited, while
at depth nutrient concentrations are high and photosynthesis is light-limited
\cite{Dugdale1967,Hodges2004}. At the depth of cross-over between these
conditions a DCM can develop \cite{Steele_Yentsch1960,Beckmann2007,Cullen2015}
and the consumption of nutrients by phytoplankton acts to fix this DCM at a
given depth.







Figure~\ref{fig:depth} shows Wasserstein distance and RMSE comparisons between Chlorophyll depth
profiles from two data sources---{\em in situ} and model---at 226 shared dates
between October 1988 and November 2016. Panel A shows an example of a single
Chlorophyll depth profile for the two data sources (for 2014-09-15), while all
226 depth profiles for each data source are shown in Panel B. For each
comparison (i.e. each common date), we also record an estimate of the DCM,
measured by the depth at which the maximum concentration of Chlorophyll
occurs. Panel C shows linear regressions of Wasserstein distance and RMSE on the estimated
difference in DCM between the two data sources. The higher $R^2$ of the left
panel of Figure~\ref{fig:depth}C suggests that Wasserstein distance is more effective than RMSE at
capturing the observed difference in DCM. Additionally,
Figure~S\ref{fig:depth_mass_transfer} shows that the most prominent movement
across depth---pooled across all comparisons made---is from approximately 96
meters in the {\em in situ} data, to 140 meters in model data. This indicates that in
aggregate, there is a depth-wise mismatch in the DCM between the two data
sources. Wasserstein distance uncovers the spatial mismatch without the additional step of
isolating the DCM.




\section{Conclusion}

We have demonstrated through a series of examples how Wasserstein distance can
be a useful tool for oceanographers performing the common task of comparing
scalar fields in the ocean. Our analyses focused on two time-varying Chlorophyll
datasets in the Pacific Ocean---a map defined over a longitude-latitude box in
the North Pacific and a depth profile at Station ALOHA. In several examples, we
found that Wasserstein distance was able to capture differences in seasonality, distribution shifts, and other
scientifically-relevant factors 
in ways that a pixel-wise difference could not.
For example, in the depth profile analysis, Wasserstein distance could more closely track the
changes in the deep Chlorophyll maximum than RMSE.
A further advantage over RMSE that we did not demonstrate in our examples is
that Wasserstein distance does not require the two sources to be defined on identical sets of
spatial cells.

Our Wasserstein distance-based analysis also suggested that the differences in Chlorophyll data from the model
and remote sensing observations can sometimes be larger than the
within-source seasonal variability.
The optimal transport maps that are generated in the computation of Wasserstein distance allowed
us to understand that this difference was driven by a seasonally varying set of
global-scale probability mass shifts. We also found that a key feature of these
two data sources---the estimated boundary between the subpolar zone and the
subtropical gyre---are much more similar in this region than the original
Chlorophyll maps. Analysis of Wasserstein distance on remote sensing data (further analyzed with
a linear regression with customized covariates) also helped reveal a long-term
change from 1998 to 2006 that is not present in the model data. This suggests
the usefulness of Wasserstein distance for examining spatial data over time within a single
source. Current studies often establish long-term trend terms of changes in
magnitude; Wasserstein distance detects changes in \textit{patterns}, which may help detect
long-term trends efficiently and with less uncertainty.


The demonstrations within this paper are just a starting point for the potential
uses of the Wasserstein distance.  We envisage this metric being used by many oceanographic data
scientists for a variety of comparisons, across a range of dimensions and
variables. One particular future development of interest would build on our application of
Wasserstein distance to province boundaries with exploration of this technique for more complex
applications than the single horizontal TZCF boundary demonstrated here.
Defining and testing provinces (``biomes'') in the ocean is an active area of
research \cite{Wust2020b,Sonnewald2020b}, and we believe that Wasserstein distances can provide a flexible tool to compare competing definitions of biomes.

As demonstrated in our examples, Wasserstein distance is particularly useful for model-data
comparison because models can struggle to get the physical location of some key
features in the ocean, such as the Gulf Stream. A pixel-wise comparison will
measure the magnitude of difference at rigid locations, while Wasserstein distance will focus on
the pattern change and appropriately measure this discrepancy in the
longitude-latitude space.


Further, the regression analysis in
Section~3a\ref{sec:lonlat-nonclim} suggests Wasserstein distance as a powerful tool to examine
\textit{temporal} trends in patterns rather than in magnitudes. This shows Wasserstein distance
goes far beyond simple model-data comparison, and can be useful for
analyzing spatial fields of ocean physical, biogeochemical as well as optical quantities over time.



Developing computational improvements will be important to allow for full global ocean comparisons. One
simple extension is to only allow local transports, by directly modifying the
base distances. Handling this sparser structured base distance effectively---by
building specialized software---may be an important
practicality. Faster approximations to optimal transport are popular in computer
science and machine learning applications, and can also be adopted when
analyzing ocean data.

Another methodological extension is to consider optimal transport with unequal
masses \cite{unbalanced-emd}, a natural scenario when dealing with
physical quantities in the ocean. Normalizing such data prior to analysis discards
a potentially important piece of information, which is the total amount of mass prior to normalization. When the data in a few bins are very large, the normalization can unduly flatten the probability mass in other bins.
An interesting future direction is to allow optimal
transport to borrow from physical transport to become more physically
realistic. Optimal transport
is not to be
confused with physical transport of the underlying quantity in the ocean. Instead, optimal
transport can be thought of as an alternative measure of distance that measures
pattern shifts in the space of the data. 
Nonetheless, making the optimal transport more physically constrained could be a beneficial future direction.  To do so, one could adjust the base distance $d_{ij}$ to account for factors such as natural boundaries in the ocean (e.g. two clear bodies
of water that do not mix) or ocean currents that prevent or promote movement in
certain directions. For example, by simulating Lagrangian drifts of particles under known
currents one might be able to form a more oceanographically relevant base distance that is then inputted into the Wasserstein distance calculation.

\enlargethispage{20pt}

\vspace{10mm}

\noindent {\bf Data and code.} {\it Available in \href{https://github.com/sangwon-hyun/omd}{https://github.com/sangwon-hyun/omd}.}\\

\noindent {\bf Author contributions.} {\it BJ and GF contributed to data curation. SH, AM, CLF, and JB contributed to formal analysis. CM and JB contributed to funding acquisition. SH, AM, CLF, CM, and JB contributed to investigation and methodology. SH, CLF, and JB contributed to project administration. SH, AM, and BJ contributed to software. JB and CM contributed to supervision.
SH, AM, and GF contributed to validation and visualization. All authors contributed to writing (both original draft, review and editing)}\\

\noindent {\bf Competing interests.} {\it We have no competing interests.}\\

\noindent {\bf Funding.} {\it This work was supported by grants by the Simons Collaboration on 
Computational Biogeochemical Modeling of Marine Ecosystems/CBIOMES 
(Grant ID: 549939 to JB; 827829 and 553242 to CLF; 549931 to MF;// 
).
Dr. Jacob Bien was also supported in part by NIH Grant R01GM123993 and NSF CAREER Award DMS-1653017. Thomas Jackson was also supported by the National Centre for Earth Observations of the UK. 
M-F Racault was also partially funded by the ``Frontiers of instability in marine ecosystems and carbon export (Marine Frontiers) [NE/V011103/1]''.}\\

\noindent {\bf Acknowledgements.} {\it The authors acknowledge the Center for Advanced Research
  Computing (CARC) at the University of Southern California for providing
  computing resources that have contributed to the research results reported
  within this publication. URL: \url{https://carc.usc.edu}.  }

\bibliographystyle{unsrtnat}
\bibliography{omd}




\newpage

\setcounter{section}{0}
\setcounter{figure}{0}
\makeatletter
\renewcommand*{\thesection}{S\arabic{section}.}
\renewcommand*{\thesubsection}{S\arabic{section}(\alph{subsection})}
\makeatother

\section{Supplement}

\subsection{Classical Multidimensional Scaling (MDS) Derivation}
\label{sec:mds-details}

The classical MDS solution $\hat Z$ is obtained by the following steps:

\begin{enumerate}
\item Calculate
  the matrix of squared distances, $D_2$, as $(D_2)_{ij} = D_{i,j}^2$.
\item Take the eigen-decomposition of the doubly centered Gram matrix
  $B = -\frac{1}{2} C D_2 C = V \Lambda V^T$, where $C = I_N - N^{-1} 1_N 1_N^T$ is the centering matrix ($I_N$ is an
  $N \times N$ identity matrix and $1_N$ is a vector of ones).
\item Take $\Lambda_{1:2}$, the $2 \times 2$ diagonal matrix of the first two
  positive eigenvalues of $B$. (Note, classical MDS is only possible when at least two positive eigenvalues exist). Take $V_{1:2}$ to be the two-column matrix of the
  corresponding two eigenvectors of $B$.
\end{enumerate}
Then, the solution is $\hat Z = D_{1:2}\Lambda_{1:2}^{1/2} V_{1:2}^T$ is a $2$-column matrix whose rows are the desired
  $2$-dimensional coordinates.

\subsection{Distance Matrix and MDS of All Monthly Data (1998-2006)}
\label{sec:nonclim-mds}

\begin{figure}[htbp!]
  \centering
  \includegraphics[width = \linewidth]{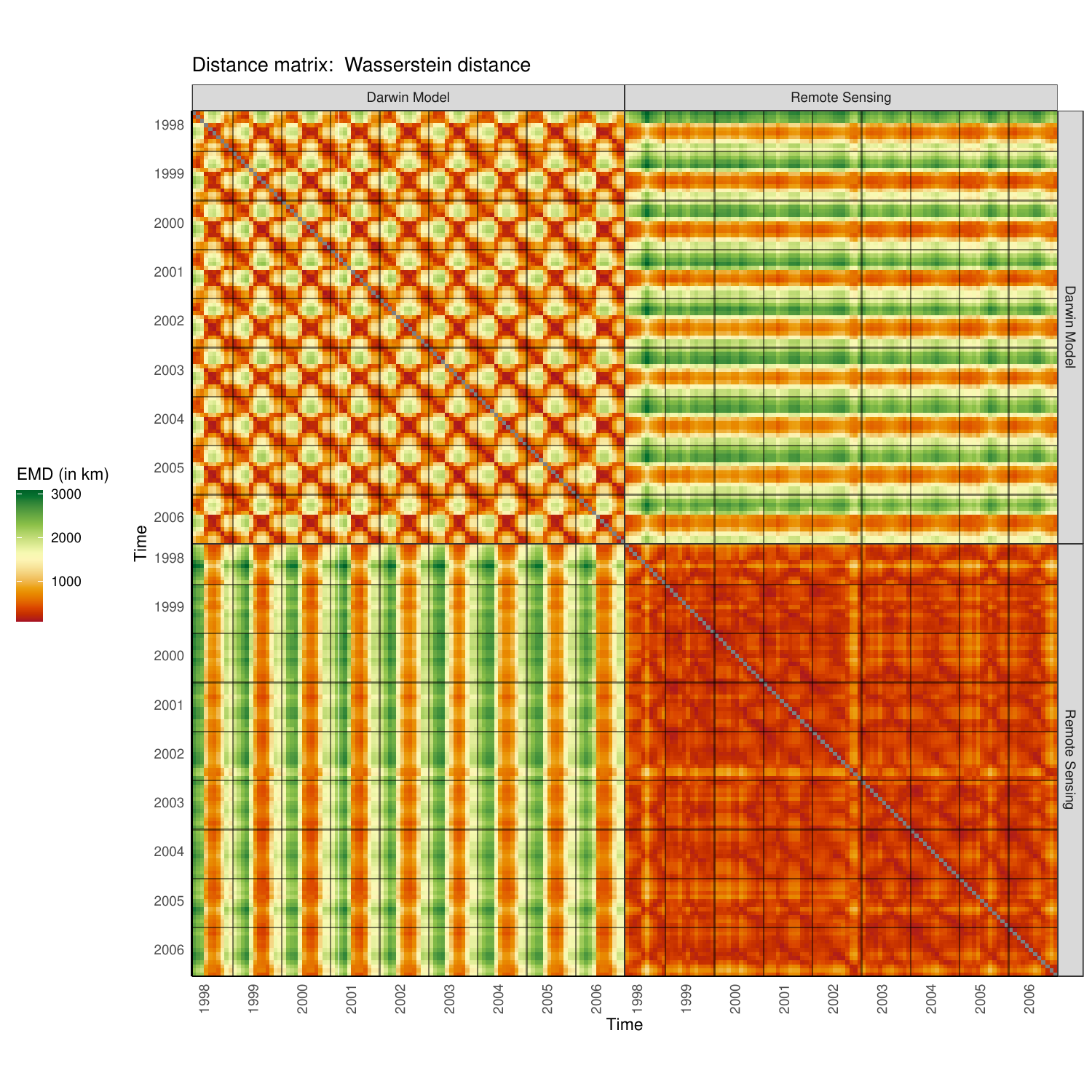}
  \caption{\it All pairwise Wasserstein distances from all monthly maps between 1998 and 2006, from
    the two data sources (Darwin and remote-sensing), as a $24 \times 24$ distance
    matrix. Figure~S\ref{fig:nonclim-mds} shows a classical MDS
    performed on this data.}
  \label{fig:nonclim-distmat}
\end{figure}

\begin{figure}[htbp!]
  \centering
  \includegraphics[width = \linewidth]{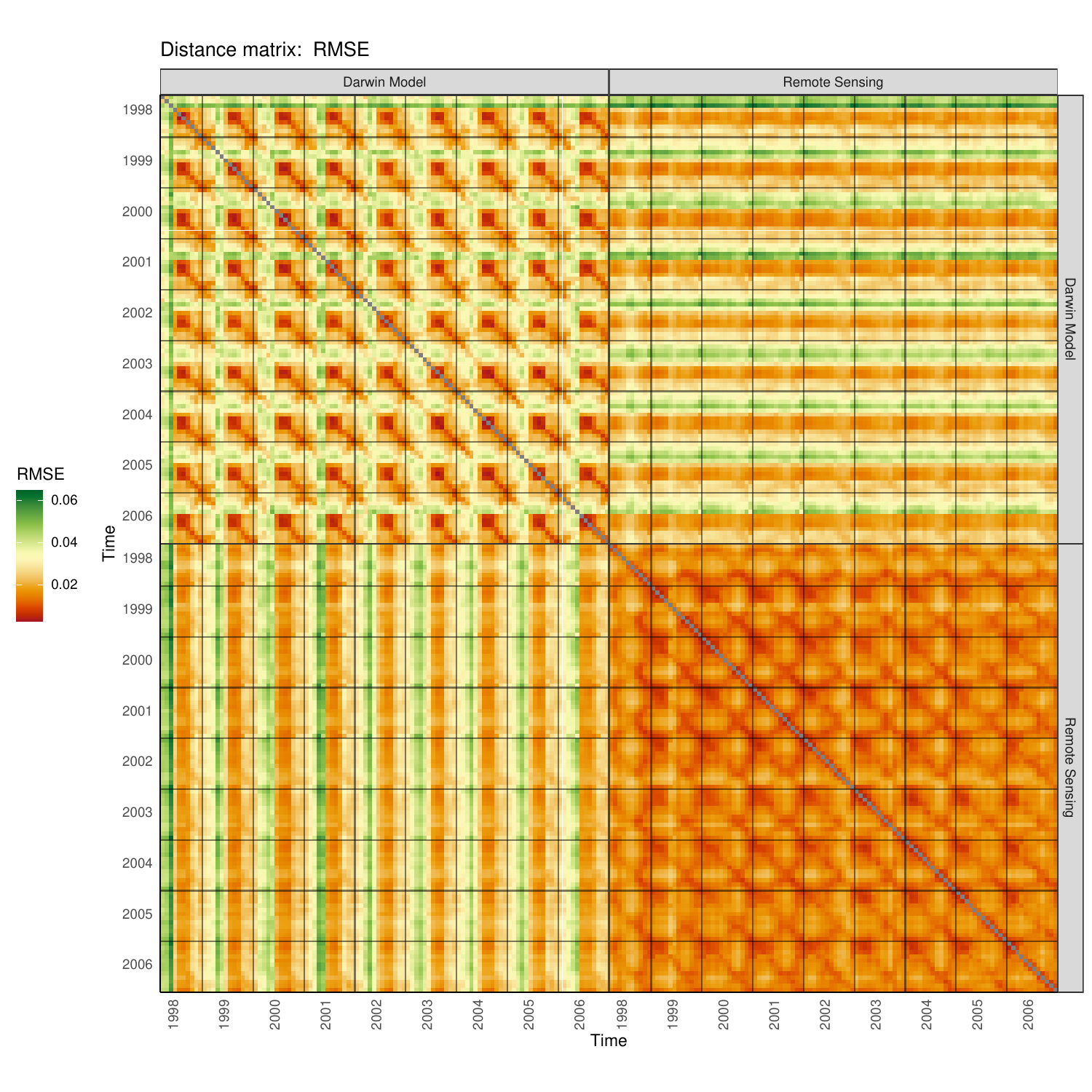}
  \caption{\it Showing RMSE distance matrix (analogous to
    Figure~S\ref{fig:nonclim-distmat}).}
  \label{fig:nonclim-distmat-rmse}
\end{figure}

\paragraph{Distance matrix.} Continuing from Section~3(a)\ref{sec:lonlat-nonclim},
the entire Wasserstein pairwise distance matrix in all months in 1998 through 2020 from
two sources (remote sensing and Darwin) are shown in Figure~S\ref{fig:nonclim-distmat}. The counterpart RMSE distance matrix is in
Figure~S\ref{fig:nonclim-distmat-rmse}.

\begin{figure}[htb!]
  \centering
  \includegraphics[width = \linewidth, trim={0 3cm 0 4.5cm},clip]{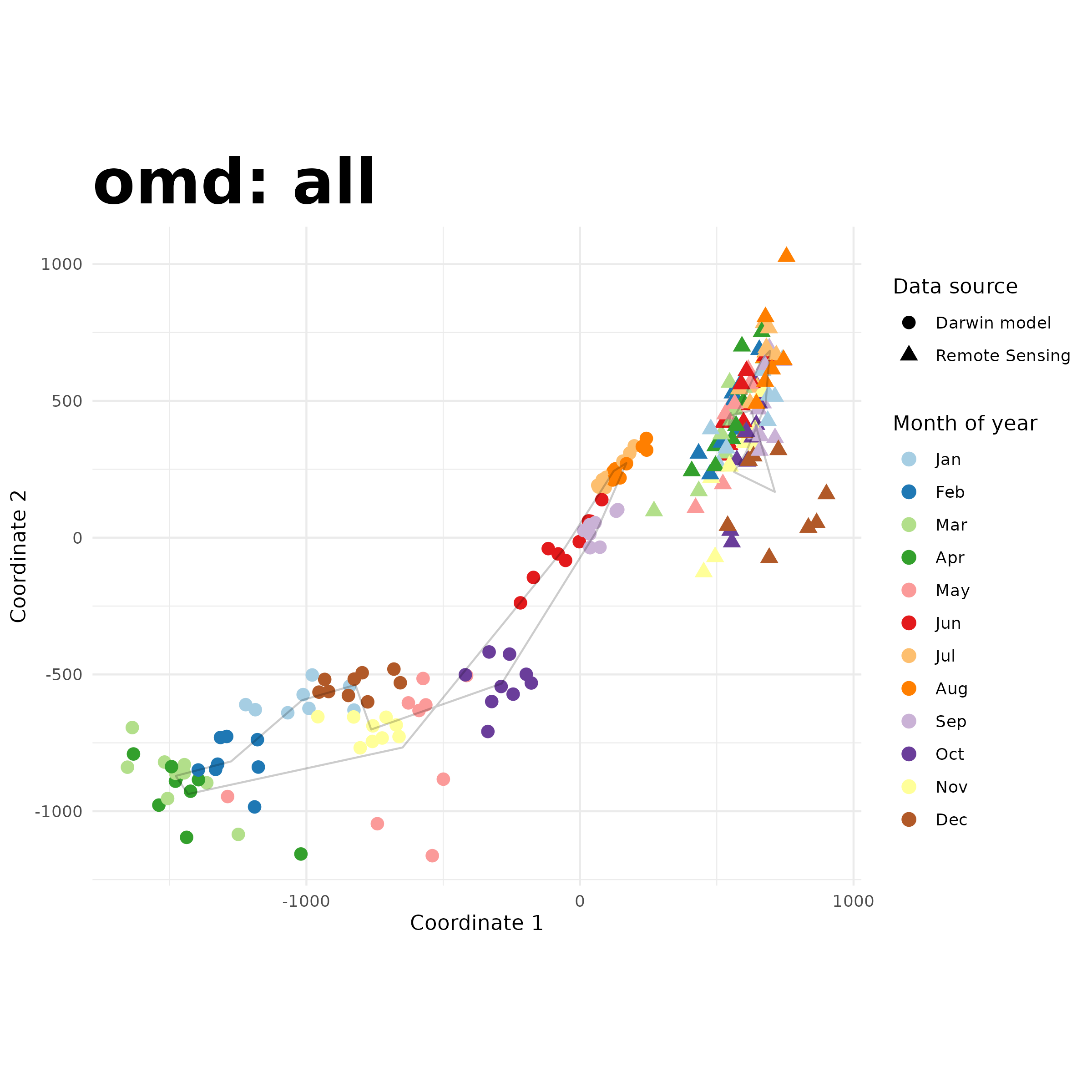} 
  \caption{\it Classical multi-dimensional scaling (MDS) performed using the distance
    matrix from Figure~S\ref{fig:nonclim-distmat}, where each month
    from each data source is plotted with two shapes---triangles for remote
    sensing and dots for Darwin model data---and colored by month of year. One
    plot for each year in 1998-2002. The same months (points with the same
    color) are clustered, which demonstrates the similarity of Chlorophyll maps
    from particular month and data source. For each data source, the grey line
    connects the average coordinate of all years within a month, and form two
    loops, which shows the seasonality in the two data
    sources. Figure~\ref{fig:nonclim-mds-separate} shows MDS plots produced
    separately from the two sources.}
  \label{fig:nonclim-mds-combined}
\end{figure}

\begin{figure}[htb!]
  \centering
  \includegraphics[width = .8\linewidth, trim={0 2.5cm 0 4.5cm},clip]{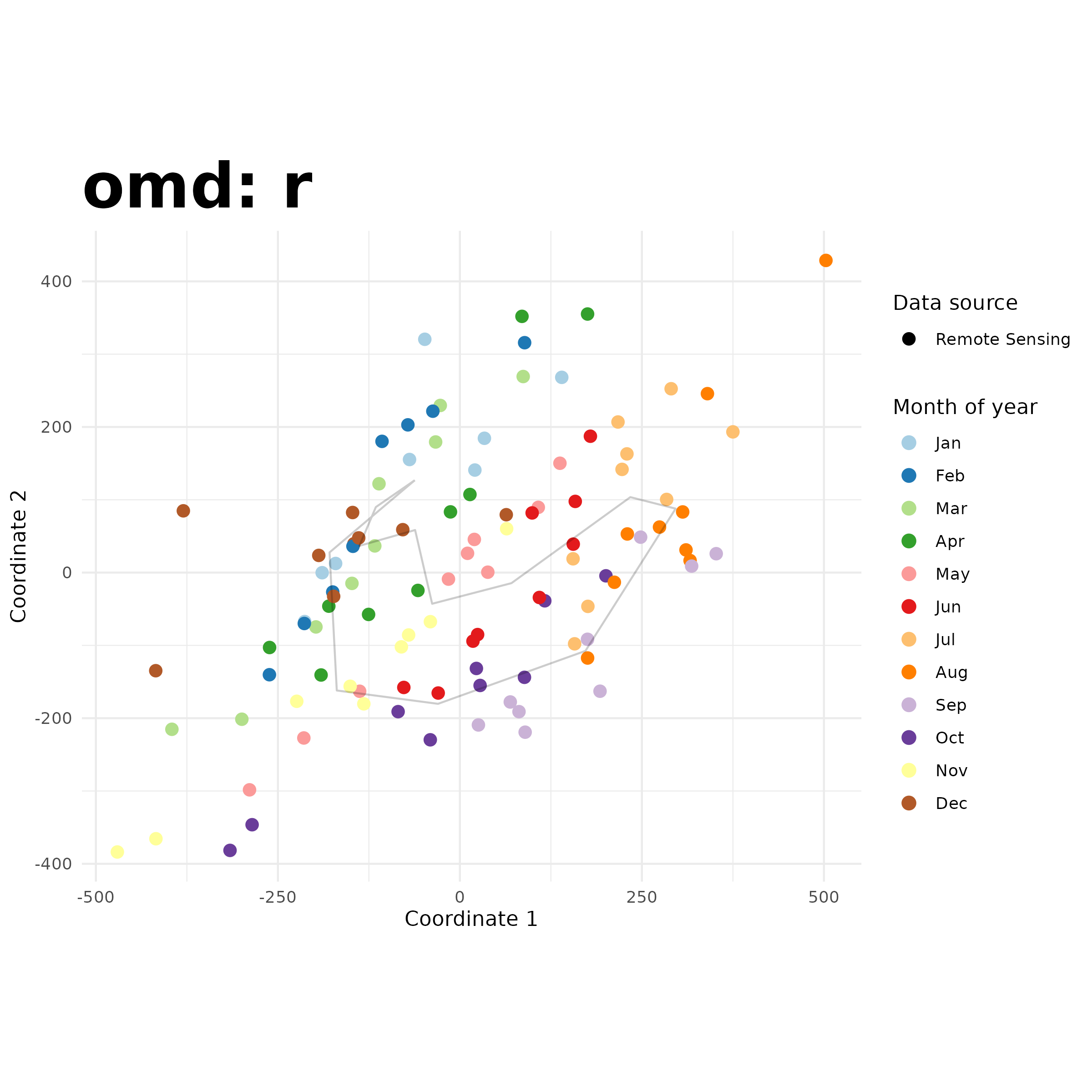} 
  \includegraphics[width = .8\linewidth, trim={0 2.5cm 0 4.5cm},clip]{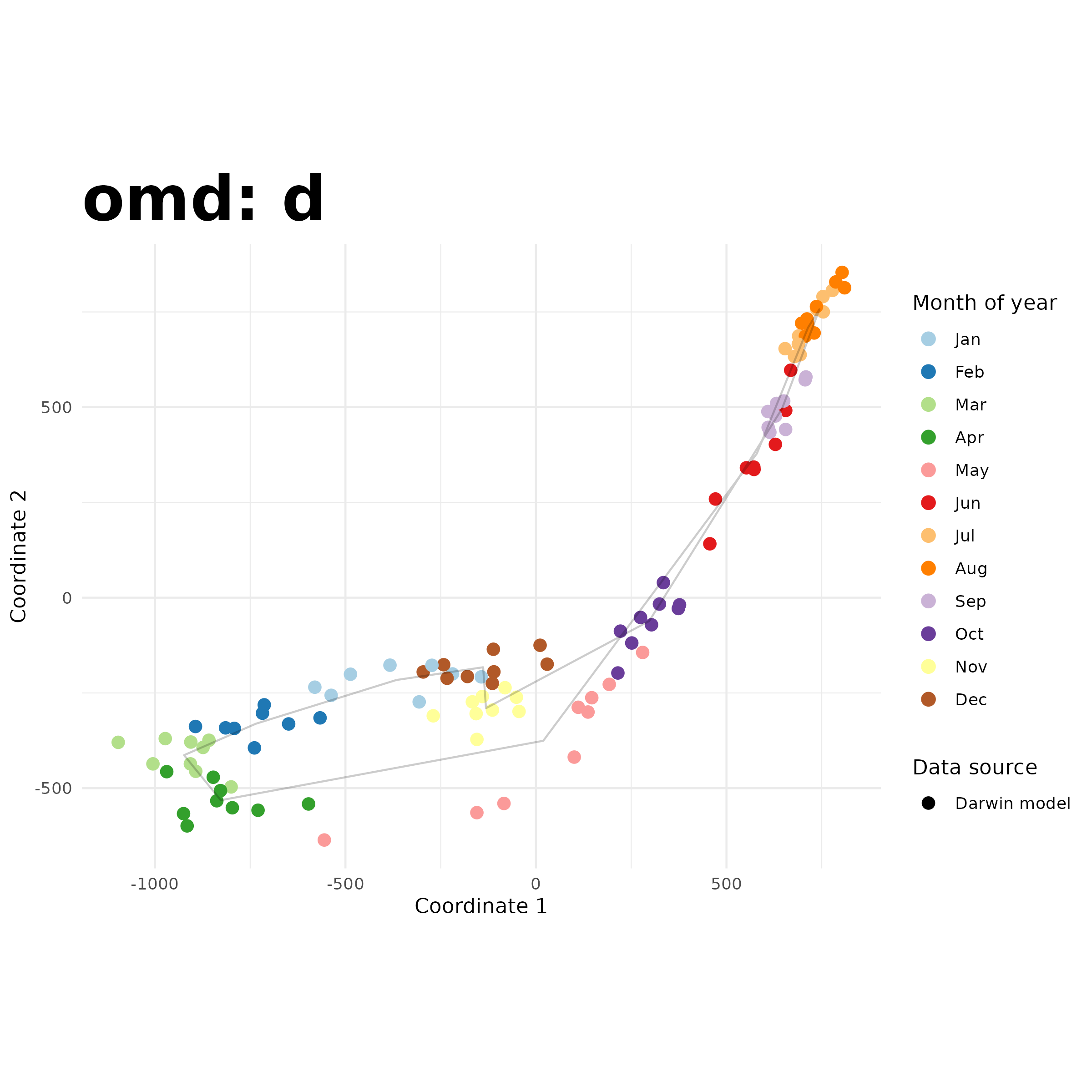} 
  \caption{\it Separate MDS plots created from the two data
    sources. (Figure~\label{fig:nonclim-mds} shows the MDS plot created from
    both sources together.) The top panel uses remote sensing, and the bottom
    panel uses Darwin data.}
  \label{fig:nonclim-mds-separate}
\end{figure}

\paragraph{MDS results.}  In the MDS plots in
Figure~S\ref{fig:nonclim-mds-combined} and S\ref{fig:nonclim-mds-separate}, we
can see clear seasonality within each source by following the two closed loops
of the grey lines that connect the average coordinates of each month from each
source.  Furthermore, we observe that the same months from the different years
cluster together---the same color points with the same shape cluster
together. As with climatology data, Darwin data has a higher variance than
remote sensing data.  The distance between the two point types (circle and
triangle) that are in the same month (same color) is far greater than the
distance within each type, supporting the same conclusion as before with
the climatology data---that the between-source variability is larger than the
within-source variability.

\subsection{Long-term Change of Chlorophyll Maps}

\begin{figure}[htb!]
  \includegraphics[width=\linewidth]{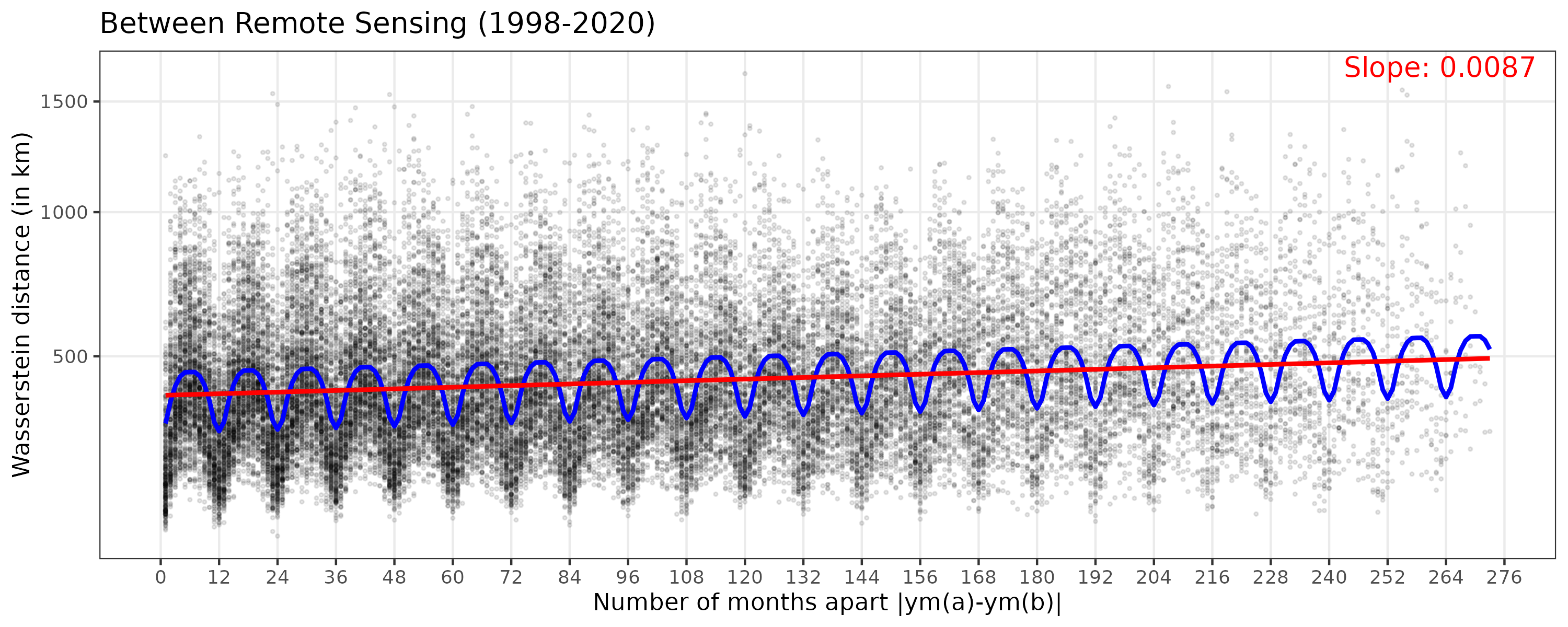}
  \caption{\it In the same style as Figure~\ref{fig:nonclim-trend}, we show the Wasserstein distance
    between Chlorophyll data from different months in a longer time range than
    before---between March 1998 to December 2020. This plot is arranged so that
    the $x$-axis shows how many months apart the two Chlorophyll maps are, and the
    $y$-axis is the Wasserstein distance (logarithmically spaced). In an estimated regression model
    \eqref{eq:regression}, the blue line shows an estimate as a function of time
    apart, and the red line shows an increasing linear trend. This additionally
    supports the conclusion from \ref{fig:nonclim-trend} that the further away
    the two months are, the larger the OMD between their Chlorophyll maps is. The slope of the red line is $0.0087$, which is about $3.7$ times larger than the slope from Darwin simulation data.
    }
  \label{fig:nonclim-trend-long}
\end{figure}

\begin{figure}[htb!]
  \includegraphics[width = \linewidth, trim={0 3cm 0 6.4cm},clip]{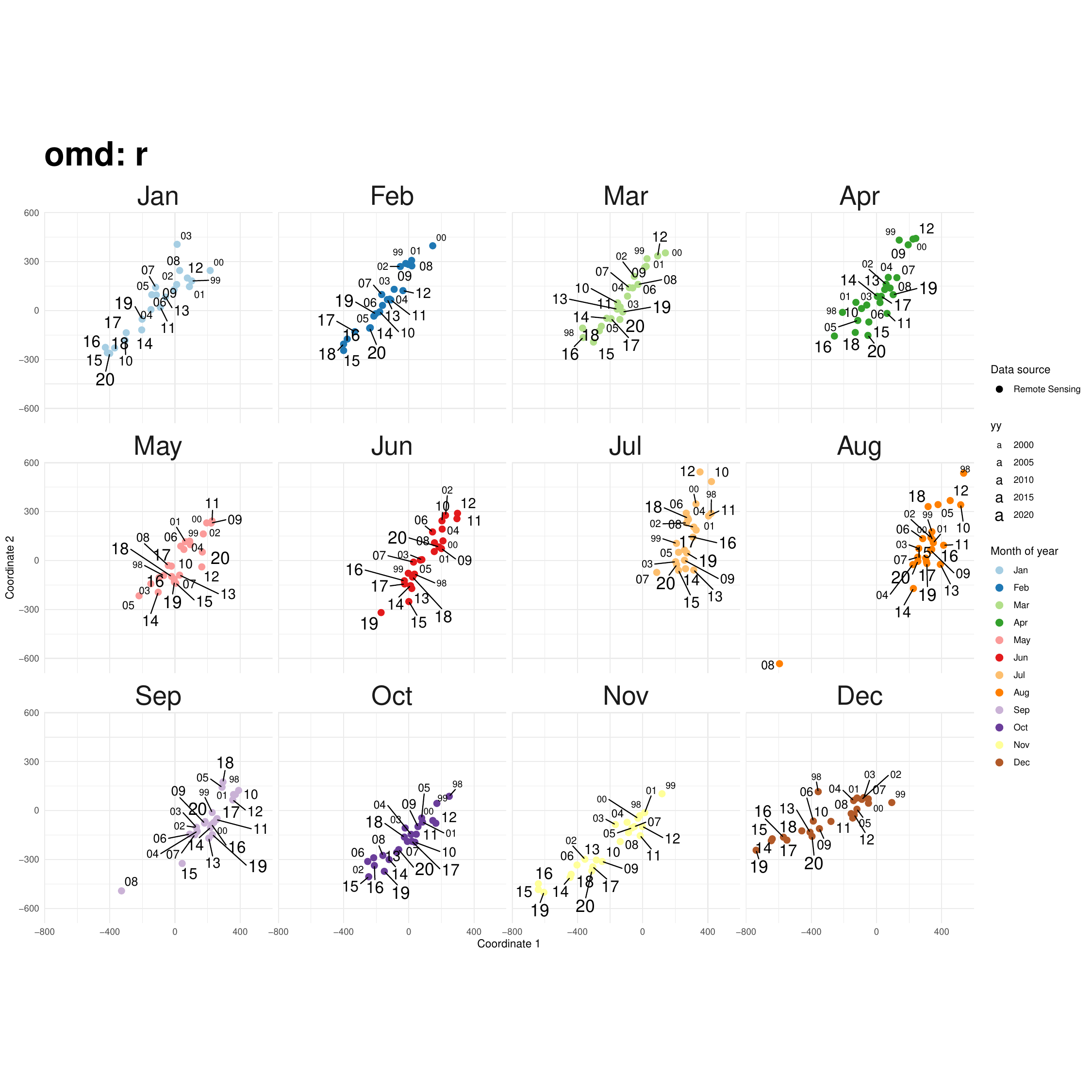}
  \caption{\it MDS of the interannualtime-resolved monthly remote sensing data in 1998-2020;
    this shows a long-term change of each calendar month. The labels show the
    last two digits of the four-digit year number. In many calendar months, the
    bottom-left of the points show later years (larger text). This shows
    additional evidence of a long-term change over time that was observed in
    Figure~S\ref{fig:nonclim-trend-long}.}
  \label{fig:nonclim-trend-long-from-mds}
\end{figure}

We find a long term temporal trend of Remote sensing Chlorophyll data in two
ways---using a specific regression model of the Wasserstein distances (Figure~S\ref{fig:nonclim-trend-long}, analyzed using the regression model in
\eqref{eq:regression}), and also using a classical MDS of remote sensing
Chlorophyll data (Figure~S\ref{fig:nonclim-trend-long-from-mds}). The latter analysis is done by
examining the MDS of the monthly remote sensing data in 1998-2020 by plotting
each month as separate panels. We can see a general drift over the years in the
following way. In each panel, each point represents a unique month in this time
period, and the point labels are the last two digits of the four-digit year
number. The size of the label is smaller for earlier years and larger for later
years. We can see that in most years, there is an overall increase along the
diagonal direction, from top-right to bottom-left.

\subsection{RMSE Instead of Wasserstein distance in Trend Plot}
\begin{figure}[htb!]
  \includegraphics[width =  \linewidth]{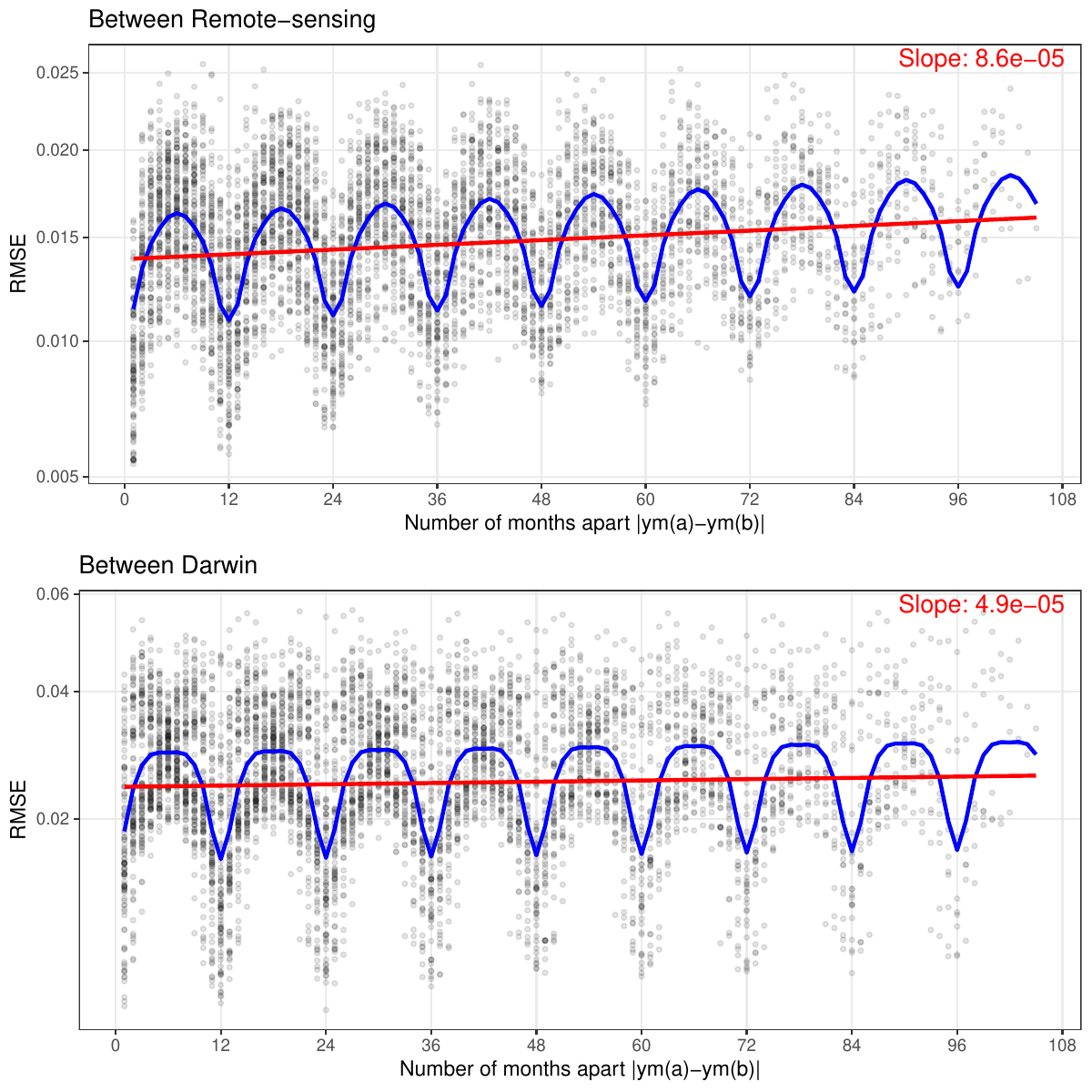}
  \caption{\it Continuing from Figure~\ref{fig:nonclim-trend} (which shows Wasserstein distance),
    this shows RMSE between Chlorophyll data from all months between March 1998
    to December 2006 (for the remote sensing (top panel) and model (bottom panel)
    data), arranged so that the $x$-axis shows how many months apart the two
    Chlorophyll maps are, and the $y$-axis is the Wasserstein distance (logarithmically
    spaced). The blue line is a regression estimate that explicitly accounts for
    seasonality (in the number of calendar months apart), and the red line is
    the long-term trend line excluding this seasonality. The slope for remote
    sensing data is positive and about 2 times larger than for Darwin data.}
  \label{fig:nonclim-trend-rmse}
\end{figure}

Figure~S\ref{fig:nonclim-trend-rmse} shows the same method used in
Figure~\ref{fig:nonclim-trend} but measured using RMSE.

\subsection{All Boundary Estimates}

\begin{figure}[htb!]
    \centering
    \includegraphics[width=\linewidth]{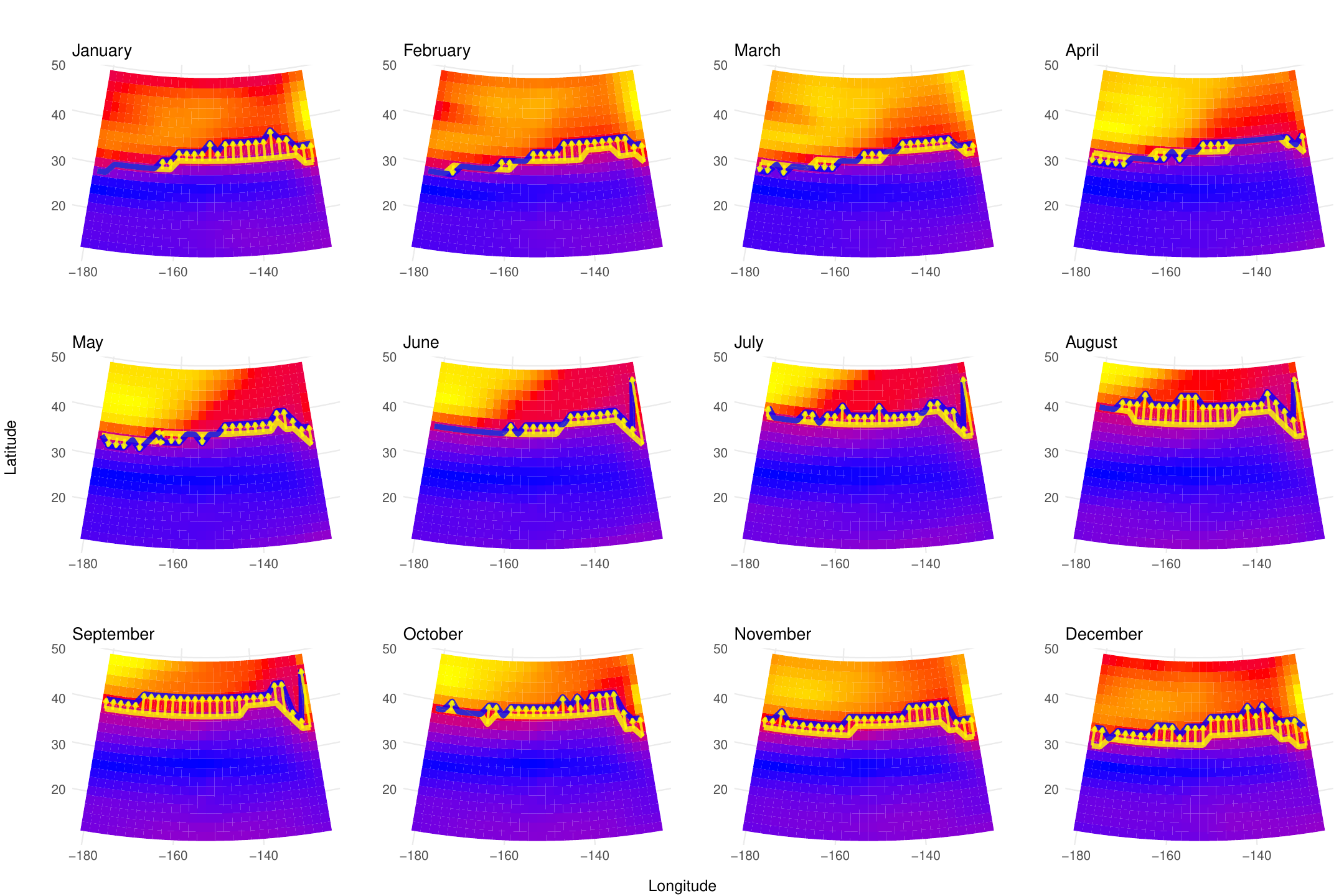}
    \caption{\it In each of the twelve months of climatology data, the twelve panels
      show province boundaries estimated from Chlorophyll maps from two sources
      (remote sensing and Darwin), shown as yellow (Darwin) and blue (remote
      sensing) lines overlaid on Darwin Chlorophyll data shown as heatmaps. The
      province boundaries were estimated using a 2-cluster K-means clustering algorithm 
      applied to vertical slices of each dataset. Then, optimal
      transports between two province boundary maps---with constant mass placed
      on the boundaries, and zero otherwise---are calculated.}
    \label{fig:boundary-all}
\end{figure}

Figure~S\ref{fig:boundary-all} shows all the boundary estimates in
all months of climatology data from the two sources (Darwin and remote sensing),
of which a subset was shown in Section~3(a)\ref{sec:boundary-omd}.

\subsection{Additional Depth Analysis}

\begin{figure}
\centering
  \includegraphics[width = 0.5\linewidth,page = 1]{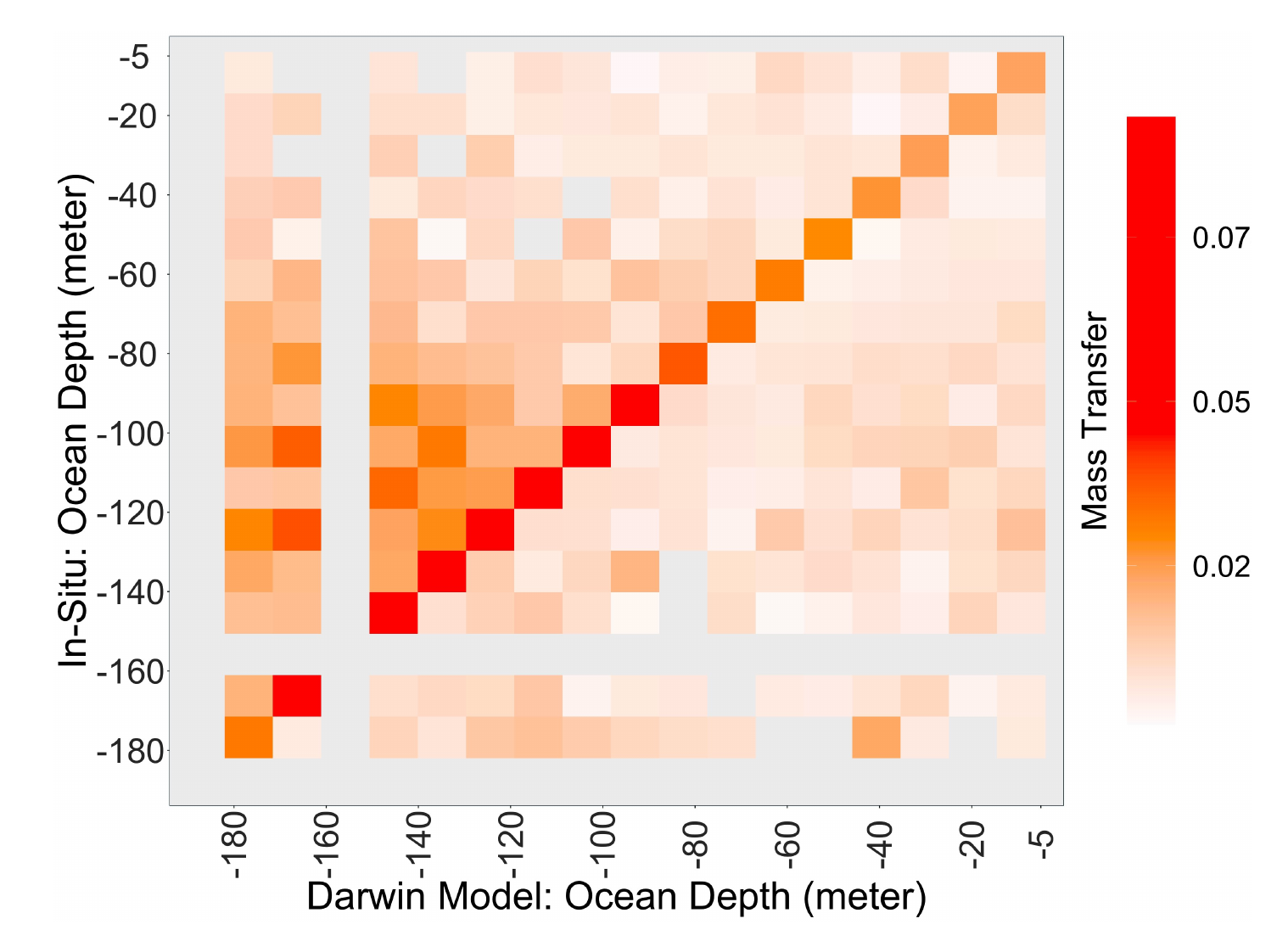}
  \caption{\it This plot reports the aggregated probability mass transferred from
    depth profiles in the Darwin model data to the remote sensing data using
    optimal transport, from all comparisons on all shared dates. See
    Figure~\ref{fig:depth} for full analysis.}
  \label{fig:depth_mass_transfer}
\end{figure}

Additional depth analysis of Chlorophyll is shown in Figure
~S\ref{fig:depth_mass_transfer}.

\end{document}